\input{style/arxiv-ba.cfg}
\documentclass[ba,linksfromyear,preprint]{imsart}
\makeatletter
   \@ifpackageloaded{natbib}{}{\usepackage{natbib}}
\makeatother
\usepackage{booktabs}

\pubyear{2015}
\volume{10}
\issue{2}
\firstpage{441}
\lastpage{478}
\doi{10.1214/14-BA913}

\startlocaldefs

\newcommand{\localscore}[2]{\ensuremath{c(#1,#2)}}
\newcommand{\varsubset}{\ensuremath{Pa}}
\newcommand{\cluster}{\ensuremath{C}}
\newcommand{\x}[1]{\ensuremath{I(#1)}}
\newcommand{\fv}[2]{\ensuremath{\x{#1 \leftarrow #2}}}
\newcommand{\gobnilp}{\textsf{GOBNILP}}
\newcommand{\scip}{\textsf{SCIP}}

\endlocaldefs

\begin{document}

\begin{frontmatter}
\title{Searching Multiregression Dynamic Models of Resting-State fMRI
Networks Using Integer Programming}
\runtitle{Searching MDM Using IPA}

\begin{aug}
\author{\fnms{Lilia} \snm{Costa}\thanksref{t1}\ead[label=e1]{liliacosta@ufba.br}},
\author{\fnms{Jim} \snm{Smith}\thanksref{t2}\ead[label=e2]{J.Q.Smith@warwick.ac.uk}},
\author{\fnms{Thomas} \snm{Nichols}\thanksref{t3}\ead[label=e3]{t.e.nichols@warwick.ac.uk}},
\author{\fnms{James} \snm{Cussens}\thanksref{t4}\ead[label=e4]{james.cussens@york.ac.uk}},
\author{\fnms{Eugene~P.}~\snm{Duff}\thanksref{t5}\ead[label=e5]{eduff@fmrib.ox.ac.uk}},
\and
\author{\fnms{Tamar R.} \snm{Makin}\thanksref{t6}\ead[label=e5]{tamar.makin@ndcn.ox.ac.uk}}

\runauthor{L. Costa, J. Smith, T. Nichols, J. Cussens, E.~P. Duff, T.~R. Makin}


\thankstext{t1}{The University of Warwick, UK; Universidade Federal da
Bahia, BR, {liliacosta@ufba.br}}
\thankstext{t2}{The University of Warwick, UK, {J.Q.Smith@warwick.ac.uk}}
\thankstext{t3}{The University of Warwick, UK, {t.e.nichols@warwick.ac.uk}}
\thankstext{t4}{The University of York, UK, {james.cussens@york.ac.uk}}
\thankstext{t5}{FMRIB Centre, Oxford University, UK, {eduff@fmrib.ox.ac.uk}}
\thankstext{t6}{FMRIB Centre, Oxford University, UK, {tamar.makin@ndcn.ox.ac.uk}}

\end{aug}

%
\begin{abstract}
A Multiregression Dynamic Model (MDM) is a class of multivariate time
series that represents various dynamic causal processes in a graphical
way. One of the advantages of this class is that, in contrast to many
other Dynamic Bayesian Networks, the hypothesised relationships
accommodate conditional conjugate inference. We demonstrate for the
first time how straightforward it is to search over all possible
connectivity networks with dynamically changing intensity of
transmission to find the Maximum a Posteriori Probability (MAP) model
within this class. This search method is made feasible by using a novel
application of an Integer Programming algorithm. The efficacy of
applying this particular class of dynamic models to this domain is
shown and more specifically the computational efficiency of a
corresponding search of 11-node Directed Acyclic Graph (DAG) model
space. We proceed to show how diagnostic methods, analogous to those
defined for static Bayesian Networks, can be used to suggest
embellishment of the model class to extend the process of model
selection. All methods are illustrated using simulated and real
resting-state functional Magnetic Resonance Imaging (fMRI) data.
\end{abstract}

%
\begin{keyword}
\kwd{Multiregression Dynamic Model}
\kwd{Bayesian Network}
\kwd{Integer Program Algorithm}
\kwd{Model Selection}
\kwd{Functional magnetic resonance imaging (fMRI)}
\end{keyword}


\end{frontmatter}


\section{Introduction}

In this paper a class of Dynamic Bayesian Network (DBN) models called
the Multiregression Dynamic Model (MDM) is applied to resting-state
functional Magnetic Resonance Imaging (fMRI) data. Functional MRI
consists of a dynamic acquisition, \emph{i.e.}\ a series of images,
which provides a time series at each volume element or voxel. These
data are indirect measurements of blood flow, which in turn are related
to neuronal activity. A traditional fMRI experiment consists of
alternating periods of active and control experimental conditions and
the purpose is to compare brain activity between two different
cognitive states (\emph{e.g.}\ remembering a list of words versus just
passively reading a list of words). In contrast, a ``resting-state"
experiment is conducted by having the subject remain in a state of
quiet repose, and the analysis focuses on understanding the pattern of
connectivity among different cerebral areas. The ultimate (and
ambitious) goal is to understand how one neural system influences
another \citep{poldrack2011handbook}. Some studies assume that the
connection strengths between different brain regions are constant.
Dynamic models have been proposed for resting-state fMRI, but they
usually estimate the temporal correlation between brain regions (rather
than the influence that one region exerts on another) or their scores
are not a closed form which complicates the process of learning the
network \citep[see \emph{e.g.}][]{chang2010time,Allen}. However,
clearly a more promising strategy would be to perform a search over a
large class of models that is rich enough to capture the dynamic
changes in the connectivity strengths that are known to exist in this
application. The Multiregression Dynamic Model (MDM) can do just this
\citep{queen1993multiregression,queen2009intervention} and in this
paper we demonstrate how it can be applied to resting fMRI.

To our knowledge, we present here the first application of Bayes factor
MDM search. As with standard BNs, the Bayes factor of the MDM can be
written in closed form, and thus the model space can be scored quickly.
However, unlike a static BN that has been applied to this domain, the
MDM models \emph{dynamic} links and so allows us to discriminate
between models that would be Markov equivalent in their static
versions. Furthermore the directionality exhibited in the MDM graph can
be associated with a causal directionality in a very natural way \citep
{queen2009intervention} which is also scientifically meaningful. Even
for the moderate number of processes needed in this application the
model space we need to search is very large; for example, a graph with
just 6 nodes has over 87 million possible BNs, and for a 7 node graph
there are over 58 billion \citep{steinsky2003enumeration}. Instead of
considering approximate search strategies, we exploit recent
technological advantages to perform a full search of the space, using a
recent algorithm for searching graphical model spaces, the Integer
Programming Algorithm \citep[IPA;][]{cussens11}. We are then able to
demonstrate that the MDM-IPA is not only a useful method for detecting
the existence of brain connectivity, but also for estimating its direction.

This paper also presents new prequential diagnostics customised to the
needs of the MDM, analogous to those originally developed for static
BNs, using the closed form of the one-step ahead predictive
distribution \citep{Cowell99}. These diagnostic methods are essential
because it is well known that Bayes factor model selection methods can
break down whenever no representative in the considered model class
fits the data well. It is therefore extremely important to check that
selected models are consistent with the observed series. We propose a
strategy of using the MDM-IPA to initially search across a class of
simple linear MDMs which are time homogeneous, linear and have no
change points.

We then check the best model using these new prequential diagnostics.
In practice we have found that the linear MDMs usually perform well for
most nodes receiving inputs from other nodes. However, when diagnostics
discover a discrepancy of fit, the MDM class is sufficiently expressive
for it to be embellished to accommodate other anomalous features. For
example, it is possible to include time dependent error variances,
change points, interaction terms in the regression and so on, to better
reflect the underlying model and refine the analysis. Often, even after
such embellishment, the model still stays within a conditionally
conjugate class. Therefore if our diagnostics identify a serious
deviation from the highest scoring simple MDM, we can adapt this model
and its high scoring neighbours with features explaining the
deviations. The model selection process using Bayes factors can then be
reapplied to discover models that describe the process even better. In
this way, we can iteratively augment the fitted model and its highest
scoring competitors with embellishments until the search class
accommodates the main features observed in the dynamic processes well.
This is one advantage of adopting a fully Bayesian methodology to
perform this analysis. Standard Bayesian diagnostics can be adapted to
provide guidance in checking and where necessary to guide the
modification of the model class.

The remainder of this paper is structured as follows. Section 2
provides a review of methods used to estimate connectivity and Section
3 shows the class of MDMs and a comparison between the MDM and these
other methods. Section 4 then describes the MDM-IPA used to learn a
network, and its performance is investigated using synthetic data.
Section 5 gives diagnostic statistics for an MDM whilst its application
to real fMRI data is shown in Section 6. Directions for future work are
given in Section 7.

\section{A Review of Some Methods for Discovering Connectivity}
\label{other methods}

Causal inference provides a set of statistical methods to allow us to
tentatively move ``from association to causation" \citep
{pearl2009causal}, or ``from \emph{functional} to \emph{effective
connectivity}" in neuroimaging terminology. In the study of functional
integration, which considers how different parts of the brain work
together to yield behaviour and cognition, a distinction is made
between \emph{functional connectivity} and \emph{effective
connectivity}. The former is defined as correlation or statistical
dependence among the measurements of neuronal activity of different
areas, whilst the effective connectivity can be seen as a temporal
dependence between brain areas and therefore it may be defined as
dynamic \citep[activity-dependent;][]{friston2011functional}.

\begin{figure}[!htb]
\centering
\includegraphics{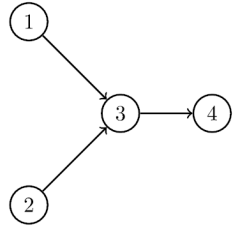}
\caption{A graphical structure considering $4$ nodes.}
\label{fig5new}
\end{figure}

A causal analysis is often represented using a directed graph where the
tail of an arrow identifies a cause and the head its effect (see \emph
{e.g.} Figure \ref{fig5new}). Graphical models have been developed in
order to define and discover putative causal links between variables
\citep{lauritzen1996graphical,pearl2000causality,spirtes2000causation}.
In this approach, the causal concepts are
expressed in term of conditional independence among variables,
expressed through a Bayesian Network (BN; see Section \ref{mdm}), which
are then extrapolated into a controlled domain \citep
{korb2003bayesian}. In such a graph, when there is a directed edge from
one node to another, the former is called a \emph{parent} while the
latter is a \emph{child}.

%

A simplified approach for estimating connectivity was proposed by
\citet{patel2006bayesian}. This is based on a comparison between conditional
and marginal probability of elevated activity. Firstly the time series
variables were dichotomized according to a certain threshold used to
indicate whether\vadjust{\eject} an elevated activity appeared at a given time point.
For instance, \citet{smith2011network} used a range of thresholds for
each time series as 10th, 25th, 50th, 75th and 90th percentile. \citet
{patel2006bayesian} then calculated a measure $\kappa$ which compares
the marginal probability that one region is active with conditional
probability given that the other region is also active. In mathematical
terms, $\kappa_{ij}$ measures the distance between $P(Y^*(i) | Y^*(j))$
and $P(Y^*(i))$, and the distance between $P(Y^*(j) | Y^*(i))$ and
$P(Y^*(j))$, where $Y^*(i)$ is a dichotomized variable that represents
whether region $i$ is active. This measure is found for each pair of
brain areas. When $\kappa_{ij} = 0$, the conditional and marginal
probabilities are the same and, in this sense, it can be concluded that
regions $i$ and $j$ are not connected.

Finally, when two particular brain regions are connected (\emph{i.e.}
$\kappa_{ij} \ne0$), measure $\tau_{ij}$ is calculated based on the
ratio of the marginal probabilities of each region, $Y^*(i)$ and
$Y^*(j)$. When $\tau_{ij}>0$, the region $i$ is ascendant to the region
$j$ whilst the negative value of this measure means that the region $j$
is ascendant to the former region. By definition, the node $j$ is
called ascendant to node $i$, if the marginal activation probability of
the former node is larger than that of node $i$.

There are many existing and useful alternate classes of model which
investigate effective connectivity. For instance a dynamic version of
the BN is the Dynamic Bayesian Network (DBN), which takes account of
the dynamic nature of a process, containing certain hypotheses about
the estimation of effective connectivity and embodies a particular type
of \emph{Granger causality} \citep{granger1969investigating}.

Granger causal hypotheses have recently been expressed in a state space
form. \citet{havlicek2010dynamic} developed a dynamic version of the
multivariate autoregressive model, using a time-varying connectivity
matrix $\mathbf{A}_l(t)$. They write
\begin{eqnarray*}
\mathbf{Y}_t &=& \sum_{l=1}^L \mathbf{A}_{l}(t) \mathbf{Y}_{t-l} +
\mathbf{v}_t, \\
\mathbf{a}_t &=& \mathbf{a}_{t-1} + \textbf{w}_t \text{, ~~~~~~~~~~}
\textbf{w}_t \sim\mathcal{N} (\mathbf{0},\textbf{W}_{t}),
\end{eqnarray*}
where $L$ is the model order; $\mathbf{A}_{l}(t)$ is the $n \times n$
matrix that represents the connectivity between the past at lag $l$ and
the current observation variables, $\mathbf{Y}_t$, and $n$ is the
dimensionality of the observable, \emph{i.e.}\ the number of nodes in
the network; $t=1, \ldots, T$; $\mathbf{a}_t=\mathrm{vec}([\mathbf
{A}_1(t), \ldots, \mathbf{A}_L(t)]')$; and $\mathbf{v}_t$ is the
$n$-dimensional white Gaussian\vadjust{\eject} error with zero-mean and variance $V$
whilst $\mathbf{w}_t$ is innovation at time $t$ with the state variance
$\mathbf{W}_t$.\ These dynamic autoregressive models allow cyclic
dependences, but are very sensitive to the particular sampling rate.
Also, model selection over the full model space is very complex because
the dimension parameter space grows exponentially with maximal AR lag.

Classes like this one that directly model Granger causality have also
received severe criticism when applied to the fMRI datasets \citep
{chang2008mapping,david2008identifying,valdes2011effective,smith2012danger}.
In fact, \citet{smith2011network} discovered that
lag-based approaches like these do not perform well at identifying
connections for fMRI data, albeit only under the assumption of static
connectivity strength.

Other much more sophisticated classes of state space models have
recently been developed to model effective connectivity. These include
the Linear Dynamic System \citep[LDS;][]{smith2010identification,smith2011effective}
and the Bilinear Dynamic System \citep
[BDS;][]{penny2005bilinear,ryali2011multivariate}. \citet
{smith2011effective} define the LDS as
\begin{eqnarray*}
\mathbf{Y}_t &=& \boldsymbol{\beta}\boldsymbol{\Phi}\mathbf{s}^{t/ \{
t-L\}}+ \mathbf{v}_t\text{, ~~~~~} \mathbf{v}_t \sim\mathcal{N}(\mathbf
{0},\mathbf{V});\\
\mathbf{s}_t &=& \textbf{A}_{u_t}\mathbf{s}_{t-1} + \textbf
{D}_{u_t}\mathbf{h}_t + \textbf{w}_t \text{, ~~~~~~~~~~} \textbf{w}_t
\sim\mathcal{N} (\mathbf{0},\textbf{W}_{u_t});
\end{eqnarray*}
where the observed fMRI signal ($\mathbf{Y}_t$) is written as a
function of the parameter $\boldsymbol{\beta}$ that represents the
weight of a known convolution matrix $\boldsymbol{\Phi}$, and the past
at lag $L$ of the latent variables, \emph{i.e.} the quasi-neural level
variables, $\mathbf{s}^{t/ \{t-L\}} = (\mathbf{s}_{t-L}', \ldots,
\mathbf{s}_t')'$, being $\mathbf{s}_t = (s_t(1), \ldots, s_t(n))'$. The
additive white Gaussian error is $\mathbf{v}_t$. The matrix $\mathbf
{A}_{u_t}$ represents the relationships among the latent variables, and
is therefore responsible for estimating the effective connectivities
whilst the matrix $\mathbf{D}_{u_t}$ is the set of regression
coefficients of driving inputs ($h_t$) on the latent variables; $u_t$
indexes the different connectivity states over the duration of the
experiment. In a BDS, $\textbf{A}_{u_t}=\mathbf{A}+\mathbf{B}\boldsymbol
{\Lambda}_t$, where $\mathbf{A}$ indicates the interactions among
latent variables without considering the influence of the experimental
condition whilst the $\mathbf{B}$ represents the connections in the
presence of modulatory inputs ($\boldsymbol{\Lambda}_t$).

One aspect of dynamic models LDS and BDS is that effective connectivity
is estimated by the interaction between the quasi-neural level
variables (rather than the observed variables). Moreover, these models
write the observed fMRI signals as a function of a convolution matrix
$\boldsymbol{\Phi}$. Methods that do not consider these two features,
\emph{i.e.} the interaction between latent variables and the
convolution matrix, nevertheless appear to correctly identify the
effective connectivity in a synthetic dataset, which was obtained under
these assumptions. For instance, \citet{smith2011network} compared
different connectivity estimation approaches based on the Dynamic
Causal Modelling (DCM) synthetic dataset (see Section \ref{IPA}). They
concluded that BNs were one of the more successful methods for
detecting network connections, although BN models had difficulty
estimating connection directionality. Another example is the study of
\citet{ryali2011multivariate}. Here they simulated data based on a BDS
model discovering that the Granger causal analysis and the BDS perform
comparably well when there are no modulatory inputs. This is the case
for resting-state data. Thus, according to the analysis of \citet
{ryali2011multivariate}, there were no significant differences in the
estimation of effective connectivity when the interaction was among
observation or latent variables, or when the model included the
convolution matrix or not.

Another popular approach in the neuroscience literature estimates
effective connectivity using DCM \citep{friston2003dynamic,stephan2008nonlinear}.
The models are quite complex in structure and
aim to capture specific scientific hypotheses. The deterministic DCM
assumes that the latent variables are completely determined by the
model, \emph{i.e.}\ the state variance is considered to be zero. As
this version of DCM does not consider the influence of random
fluctuation in neuronal activity, it cannot be used for resting-state
connectivity \citep{penny2005bilinear,smith2011effective}. More
recently a stochastic DCM has been developed that addresses this
problem \citep[\emph{e.g.}][]{daunizeau2009variational,li2011generalised}.
Both versions of the DCM depend on a nonlinear
biophysical ``Balloon model'', making the inference process quite
complex and unfeasible with more than but a few nodes \citep
{stephan2010ten,poldrack2011handbook}. Furthermore, several authors
have criticised the use of the Balloon model as speculative \citep
{roebroeck2011identification,ryali2011multivariate}, as there are
alternate plausible models for the fMRI response.

Synchronization phenomena are also studied to investigate the
communication between different brain areas \citep
{quiroga2002performance,pereda2005nonlinear,dauwels2010comparative}.
\citet{arnhold1999robust} proposed two non-linear measures to estimate
the interdependence between two particular regions as
\begin{eqnarray*}
S_{ij} &=& \frac{1}{T} \sum_{t=1}^T \frac{R_t^{(k)}(\mathbf{Y}(i))}{
R_t^{(k)}(\mathbf{Y}(i)|\mathbf{Y}(j))}\text{ and}\\
H_{ij} &=& \frac{1}{T} \sum_{t=1}^T \log\frac{R_t(\mathbf{Y}(i))}{
R_t^{(k)}(\mathbf{Y}(i)|\mathbf{Y}(j))},\\
\end{eqnarray*}
where $T$ is the sample size, $R_t^{(k)}(\mathbf{Y}(i))$ is the mean
squared Euclidean distance to the $k$ nearest neighbours of the time
series of region $i$, $R_t^{(k)}(\mathbf{Y}(i)|\mathbf{Y}(j))$ is the
mean squared Euclidean distance between $\mathbf{Y}(i)$ and the $k$
nearest neighbours of the time series of region $j$, and
\begin{eqnarray*}
R_t(\mathbf{Y}(i)) = (1/T)\sum_{t=1}^T R_t^{(T-1)}(\mathbf{Y}(i)).
\end{eqnarray*}
In addition, \citet{quiroga2002performance} suggested a new measure as
\begin{equation*}
N_{ij} = \frac{1}{T} \sum_{t=1}^T \frac{R_t(\mathbf{Y}(i))-
R_t^{(k)}(\mathbf{Y}(i)|\mathbf{Y}(j))}{ R_t(\mathbf{Y}(i))}.
\end{equation*}
The measures $H_{ij}$ and $N_{ij}$ are more robust against noise than
$S_{ij}$, but the former is not normalised whilst $N_{ij}$ is, assuming
values between zero and one. Although these measures are defined as
being theoretically different, \citet{quiroga2002performance} found
similar results when these measures were applied to real datasets. Also
\citet{smith2011network} reported that $H_{ij}$ and $N_{ij}$ provide
similar results using synthetic data.

A Linear Non-Gaussian Acyclic Model (LiNGAM) is used to estimate
effective connectivity, based on the following assumptions: (1) data
are generated through a linear process consistent with an acyclic
graphical structure; (2) unobserved confounders are not allowed; (3)
noise variables are mutually independent and have non-Gaussian
distributions with non-zero variances \citep{shimizu2006linear}. Thus
suppose $\mathbf{Y}$ is the observed data matrix. Then LiNGAM consists
of the model:
\begin{equation*}
\mathbf{Y} = \mathbf{BY} + \mathbf{e};
\end{equation*}
where $\mathbf{B}$ is a lower triangular matrix with all zeros on the
diagonal and $\mathbf{e}$ is a residual matrix. Solving for $\mathbf
{Y}$, $\mathbf{Y} = \mathbf{Ae}$ is obtained, where $\mathbf{A} =
(\mathbf{I} - \mathbf{B})^{-1}$ and $\mathbf{I}$ is the identity
matrix. The assumption of non-Gaussianity enables the direction of
causality to be identified so that the effective connectivity can be
estimated \citep{shimizu2006linear}. Because the components of $\mathbf
{e}$ must be mutually independent and non-Gaussian, $\mathbf{A}$ can be
estimated in an identifiable way: Independent Component Analysis (ICA).

\section{The Multiregression Dynamic Model}
\label{mdm}

In this paper we propose a further class of models that can be applied
to investigate effective connectivity. The MDM \citep
{queen1993multiregression} is a graphical multivariate model for an
n-dimensional time series $Y_t(1), \ldots, Y_t(n)$. This is a
particular dynamic generalisation of the family of Gaussian BNs, and we
begin by describing this class.

~~~\\
\noindent
\emph{The Bayesian Network (BN)}

Bayesian Network models decompose the joint distribution of a set of
observables into a set of conditional distributions. BNs embody the
assumption of the Markov property, and only consider direct
dependencies that are explicitly shown via edges \citep
{korb2003bayesian}. More explicitly, in a BN with nodes represented by
the random variables $\mathbf{Y}=(Y(1), \ldots, Y(n))$, the chain rule
allows the joint density to be factorized as the product of the
distribution of the first node and transition distributions between the
following nodes, \emph{i.e.}:
\begin{eqnarray*}
p_\mathbf{y}(y(1),y(2), \ldots, y(n)) &=& p_1(y(1)) \times
p_2(y(2)|y(1))\\
& ~& \times\ldots\times p_n(y(n)|y(1), \ldots,y(n-1))\\
&=& p_1(y(1)) \times\prod_{r=2}^n p_r(y(r)|y(1), \ldots,y(r-1)).
\end{eqnarray*}
Let $Pa(r)\subseteq\{Y(1), \ldots, Y(r-1)\}$ and call $Pa(r)$ the
parents of $Y(r)$ --- those nodes connected into $Y(r)$ by a directed
edge in the BN. The Markov properties depicted in the BN state that a
node depends only on its parents. This allows us to simplify the
expression above to
\begin{eqnarray*}
p_\mathbf{y}(y(1),y(2), \ldots, y(n)) = p_1(y(1)) \times\prod_{r=2}^n
p_r(y(r)| Pa(r)).
\end{eqnarray*}
When observed variables are jointly Gaussian, the conditional
distribution of variables is defined as $(Y(r) | Pa(r), \boldsymbol
{\theta}(r), V(r)) \sim\mathcal{N}(Pa(r)' \boldsymbol{\theta
}(r),V(r))$, where $\mathcal{N}(\cdot, \cdot)$ is a Gaussian
distribution; in this context the regression coefficient $\boldsymbol
{\theta}(r)$ represents the functional connectivity strengths (except
for the intercept); and $r = 1, \ldots, n$.

~~~\\
\noindent
\emph{A Description of the MDM}

Now consider the column vector $\mathbf{Y}'_t = (Y_t(1), \ldots,
Y_t(n))$ which denotes the data from $n$ regions at time $t$. Denote
their observed values designated respectively by $\mathbf{y}'_t =
(y_t(1), \ldots, y_t(n))$. Let the time series until time $t$ for
region $r = 1, \ldots, n$ be $\mathbf{Y}^t(r)' = (Y_1(r), \ldots,
Y_t(r))$. The MDM is defined by $n$ observation equations, a system
equation and initial information \citep{queen1993multiregression}. The
observation equations specify the time-varying regression parameters of
each region on its parents. The system equation is a multivariate
autoregressive model for the evolution of time-varying regression
coefficients, and the initial information is given through a prior
density for regression coefficients. Thus the multiregression dynamic
model is specified in terms of a collection of conditional regression
dynamic linear models \citep[DLMs;][]{West}, as follows:

We write the \emph{observation equations} as
\begin{equation*}
Y_t(r) = \mathbf{F}_t(r)' \boldsymbol{\theta}_t(r) + v_t(r) \text{,
~~~~~} v_t(r) \sim\mathcal{N}(0,V_t(r));
\end{equation*}
where $r = 1, \ldots, n$; $t = 1, \ldots, T$; $\mathbf{F}_t(r)'$ is a
covariate vector with dimension $p_r$ determined by $Pa(r)$; the first
element of $\mathbf{F}_t(r)'$ is $1$, representing an intercept, and
the remaining columns are observed time series of its parents; $p_r$ is
the number of parents of node $r$ plus $1$ (for the intercept). The
$p_r$-dimensional time-varying regression coefficient $\boldsymbol
{\theta}_t(r)$ represents the effective connectivity (except for the
intercept); and $v_t(r)$ is the independent residual error with
variance $V_t(r)$. Concatenating the $n$ regression coefficients as
$\boldsymbol{\theta}_t'= (\boldsymbol{\theta}_t(1)', \ldots, \boldsymbol
{\theta}_t(n)')$ gives a vector of length $p=\sum_{r=1}^n p_r$. We next
write the \emph{system equation} as
\begin{equation*}
\boldsymbol{\theta}_t = \mathbf{G}_t \boldsymbol{\theta}_{t-1} + \mathbf
{w}_t\text{, ~~~~~~~~~~} \mathbf{w}_t \sim\mathcal{N} (\mathbf
{0},\mathbf{W}_t); \text{ ~~~~~~~~~~~~~~~~~~~~~}
\end{equation*}
where $\mathbf{G}_t = \text{blockdiag}\{\mathbf{G}_t(1), \ldots, \mathbf
{G}_t(n)\}$, each $\mathbf{G}_t(r)$ being a $p_r \times p_r$ matrix,
$\mathbf{w}_t$ are innovations for the latent regression coefficients,
and $\mathbf{W}_t = \text{blockdiag}\{\mathbf{W}_t(1), \ldots, \mathbf
{W}_t(n)\}$, each $\mathbf{W}_t(r)$ being a $p_r \times p_r$ matrix.
The error $\mathbf{w}_t$ is assumed independent of $\textbf{v}_s$ for
all $t$ and $s$; $\textbf{v}_s= (v_s(1), \ldots, v_s(n))$. For most of
the development we need only consider $\mathbf{G}_t(r)=\mathbf
{I}_{p_r}$, where $\mathbf{I}_{p_r}$ is the $p_r$-dimensional identity
matrix. Because the errors follow a Gaussian distribution and the
relationship between the observed variables and their parents is
linear, this class of model is called a linear MDM \citep
[LMDM;][]{queen2008forecast}. Notice that by setting $\mathbf
{W}_t=\mathbf{0}$ and $\mathbf{G}_t$ as the identity matrix we retrieve
a Gaussian BN as defined above, whose regression coefficients are given
prior Gaussian distributions.

For instance, suppose the graphical structure given by Figure \ref
{fig5new} in\vadjust{\eject} Section 2 above, then the model equations are written as:
\begin{eqnarray*}
\boldsymbol{\theta}_t(r) &=& \boldsymbol{\theta}_{t-1}(r) + \mathbf
{w}_t(r);\text{ }\mathbf{w}_t(r) \sim\mathcal{N}\left(\mathbf{0},
\mathbf{W}_t(r)\right);\\
Y_{t}(1) &=& \theta_t^{(1)}(1) + v_{t}(1);\\
Y_{t}(2) &=& \theta_t^{(1)}(2) + v_{t}(2);\\
Y_{t}(3) &=& \theta_t^{(1)}(3) + \theta_t^{(2)}(3) Y_{t}(1) + \theta
_t^{(3)}(3) Y_{t}(2) + v_{t}(3);\\
Y_{t}(4) &=& \theta_t^{(1)}(4) + \theta_t^{(2)}(4) Y_{t}(3) +
v_{t}(4);\text{ } v_{t}(r) \sim\mathcal{N}\left(0, V_t(r)\right),
\end{eqnarray*}
for $r=1, \dots, 4$, $p_1 = 1$, $p_2 = 1$, $p_3 = 3$ and $p_4 = 2$. The
effective connectivity strengths of this example are then $\theta
_t^{(2)}(3)$, $\theta_t^{(3)}(3)$ and $\theta_t^{(2)}(4)$.

Finally the \emph{initial information} is written as
\begin{equation*}
(\boldsymbol{\theta}_0| y_0) \sim\mathcal{N} (\mathbf{m}_0, \mathbf{C}_0);
\end{equation*}
where $\boldsymbol{\theta}_0$ expresses the prior knowledge of the
regression parameters, before observing any data, given the information
at time $t=0$, \emph{i.e.} $y_0$. The mean vector $\mathbf{m}_0$ is an
initial estimate of the parameters and $\mathbf{C}_0$ is the $p \times
p$ variance-covariance matrix. $\mathbf{C}_0$ can be defined as $\text
{blockdiag}\{\mathbf{C}_0(1), \ldots, \mathbf{C}_0(n)\}$, with each
$\mathbf{C}_0(r)$ being a $p_r$ square matrix. When the observational
variances are unknown and constant, \emph{i.e.} $V_t(r) = V(r)$ for all
$t$, by defining $\phi(r)=V(r)^{-1}$, a prior
\begin{equation*}
(\phi(r)| y_0) \sim\mathcal{G}\left(\frac{n_0(r)}{2}, \frac
{d_0(r)}{2}\right),
\end{equation*}
where $\mathcal{G}(\cdot, \cdot)$ denotes a Gamma distribution, leads
to a conjugate analysis where conditionally each component of the
marginal likelihood has a Student t distribution. In order to use this
conjugate analysis it is convenient to reparameterise the model as
$\mathbf{W}_t(r) = V(r)\mathbf{W}_t^*(r)$ and $\mathbf{C}_0(r) = V(r)
\mathbf{C}_0^*(r)$. For a fixed innovation signal matrix $\mathbf
{W}_t^*(r)$ this change implies no loss of generality \citep{West}.

This reparametrization simplifies the analysis. In particular, it
allows us to define the innovation signal matrix indirectly in terms of
a single hyperparameter for each component DLM called a \emph{discount
factor} \citep{West,petris2009dynamic}. For the particular model
selection purposes we require here, this well tested and expedient
simplification vastly reduces the dimensionality of the model class,
whilst in practice it usually loses very little in the quality of fit.
This well used technique expresses different values of $\mathbf
{W}_t^*(r)$ in terms of the loss of information in the change of
$\boldsymbol{\theta}(r)$ between times $t-1$ and $t$. More precisely,
for some $\delta(r) \in(0,1]$, the state error covariance matrix
\begin{equation*}
\mathbf{W}_t^*(r)=\frac{1-\delta(r)}{\delta(r)}\mathbf{C}_{t-1}^*(r);
\end{equation*}
where $\mathbf{C}_t(r) = V(r) \mathbf{C}_t^*(r)$ is the posterior
variance of $\boldsymbol{\theta}_t(r)$. Note that when $\delta(r)=1$,
$\mathbf{W}_t^*(r)=0\mathbf{I}_{p_r}$, there are no stochastic changes
in the state vector.

For any choice of discount factor $\delta(r)$ and any MDM the
recurrences given above provide a closed form expression for this
marginal likelihood. This means that we can estimate $\delta(r)$ simply
by maximising this marginal likelihood, performing a direct
one-dimensional optimisation over $\delta(r)$, analogous to that used
in \citet{heard2006quantitative} to complete the search algorithm. The
selected component model is then the one with the discount factor
giving the highest associated Bayes factor score, as we will see later.

The joint density over the vector of observations associated with any
MDM series can be factorized into the product of the density of the
first node and the (conditional) transition densities between the
subsequent nodes \citep{queen1993multiregression}. Moreover, the
conditional one-step forecast distribution can be written as $(Y_t(r) |
\mathbf{y}^{t-1}, Pa(r)) \sim\mathcal{T}_{n_{t-1}(r)}(f_t(r),
Q_t(r))$, where $\mathcal{T}_{n_t(r)}(\cdot, \cdot)$ is a noncentral t
distribution with $n_t(r)$ degrees of freedom and the parameters are
easily found through \emph{Kalman filter} recurrences \citep[see \emph
{e.g.}][]{West}.
The joint log predictive likelihood (LPL) can then be calculated as
\begin{eqnarray} \label{mod}
\text{LPL}(m) = \sum_{r=1}^{n} \sum_{t=1}^T \log p_{tr}(y_t(r)|\mathbf
{y}^{t-1}, Pa(r),m),
\end{eqnarray}
where $m$ denotes the current choice of model that determines the
relationship between the $n$ regions expressed graphically through the
underlying graph. The most popular Bayesian scoring method, and the one
we use here, is the Bayes factor measure \citep{jeffreys1998theory,West}.
Strictly in this context this simply uses the LPL($m$): $m_1$ is
preferred to $m_2$ if $LPL(m_1)>LPL(m_2)$.

~~~\\
\noindent
\emph{An Overview of the LMDM}

The LMDM is a composition of simpler univariate regression dynamic
linear models \citep[DLMs;][]{West}, which can model smooth changes in
the parents' effect on a given node during the period of investigation.
There are four features of this model class that are useful for this study.
\begin{enumerate}
\item Each LMDM is defined in part by a directed acyclic graph (DAG)
whose vertices are observed fMRI series at a given time. In addition,
its directed edges represent the existence of a dependence on those
contemporaneous observations that are explicitly included as regressors
to the receiving variable. In our context, therefore, these directed
edges denote the hypothesis that direct contemporaneous relationships
might exist between a variable and its parent. The directionality of
the edges can be interpreted as being `causal' in a sense that is
carefully argued in \citet{queen2009intervention};
\item Any LMDM enables a conjugate analysis. In particular its marginal
likelihood can be expressed as a product of multivariate Student $t$
distributions, as detailed above. Its closed form allows us to perform
a fast model selection;
\item Although the predictive distribution of each node given its
parents is multivariate Student t distributed, because the covariates
enter the scale function of these conditionals, the joint distribution
can be highly\vadjust{\eject} non-Gaussian \citep[see][for an example of
this]{queen1993multiregression}. The use of joint Gaussian
distributions has been criticised in the study of fMRI data;
\item The class of LMDM can be further modified to include other
features that might be necessary in a straightforward and convenient
manner: for example by adding dependence on the past of the parents
(not just the present), allowing for change points, and other
embellishments that we illustrate below.
\end{enumerate}

~~~\\
\noindent
\emph{Comparison with Other Models Used to Analyse fMRI Experiments}

The LMDM is a dynamic version of the Gaussian BN, where, unlike the
latter, the LMDM allows the strength of connectivity to change over
time. By explicitly modelling drift in the directed connection
parameters, the LMDM is able to discriminate between models whose
graphs are Markov equivalent. By definition, two network structures are
said to be Markov equivalent when they correspond to the same
assertions of conditional independence \citep{heckerman1998tutorial}.
As a result two models, indistinguishable as BN models, become distinct
when generalised into LMDMs, see an example of this in Section \ref
{diag}. The dynamic version of the BN is the DBN. This uses a Vector
Autoregression (VAR) type time series sequentially rather than the
state space employed in the LMDM which can also be used to represent
certain Granger causal hypotheses.

In the previous literature, a sliding time window has been used to
estimate the dynamic correlation among brain regions \citep
{chang2010time,Allen,leonardi2013principal}. Some methods have
investigated change points in sparse undirected graphs \citep
{cribben2012dynamic}, or in the global structure, consisting of global
chain and V dependences among three networks \citep
{zhang2013inferring}. In contrast to the LMDM, these methods study
functional connectivity, and also use sophisticated but much more
complex statistical computational algorithms rather than conditional
conjugate analyses to perform inference.

Possibly the closest family of competitive models are the dynamic
Granger causal models \citep{havlicek2010dynamic} described above.
These use Kalman filtering to obtain the posterior distribution of
effective connectivity. However, in general, the scores of these models
are not factorable and are therefore much slower to search over. The
other methods, such as LDS, BDS and DCM, we reviewed in Section 2, are
more sophisticated but also far too complicated to effectively score
quickly enough over a large model space. Consequently these are not
good candidates for use in the initial exploratory search we have in
mind here. An important difference between these methods and LMDM is
that while the dynamic of connectivity is directly estimated in LMDM,
most other models consider connectivity as static or estimate only the
different strengths of connectivity when modelling a different
experimental situation. We show in our analyses below that in practice
these strengths seem to drift in time. Of course some authors discuss
the possibility of a connectivity for each time point, $u_t=t$,
including another dynamic system for the connectivity, \emph{i.e.}
$\mathbf{A}_t = \mathbf{A}_{t-1} + \textbf{wa}_t$, where $\textbf{wa}_t
\sim\mathcal{N} (\mathbf{0},\textbf{Wa})$ \citep
{bhattacharya2006bayesian,smith2011effective}. But the inferential
techniques needed for these models are considerably more complicated
than for LMDM (\emph{e.g.} using\vadjust{\eject} the Gibbs sampler scheme) and there
are some extra assumptions they need to make, \emph{e.g.} a fixed
variance $\mathbf{Wa}$, which are not assumed in LMDM. Therefore, to
our knowledge, no other competitive class of models generates formally
justifiable scores in a closed form, and simultaneously allows
connectivity to change over time in this way.

\citet{bhattacharya2006bayesian} proposed a similar model for LDS, but
with the assumption that the observational errors are temporally
dependent. This simpler class can also be implemented as a DLM (and so
as an LMDM), as shown by \citet[ch.\ 9]{West}. Another modification in
the model assumption was suggested by \citet
{bhattacharya2011nonstationary}. These authors proposed the
autoregressive model for effective connectivity as $\mathbf{A}_t = \rho
\mathbf{A}_{t-1} + \textbf{wa}_t$, where $\rho$ is a parameter to be
estimated. So implicitly here we are using a random walk model for
effective connectivity, \emph{i.e.} $\mathbf{G}_t=\mathbf{I}_p$. It
would be possible to use $\mathbf{G}_t=\rho\mathbf{I}_p$ and a similar
procedure provided by \citet[ch.\ 4]{petris2009dynamic} to estimate the
matrix $\mathbf{G}$ within our method. However, this again gives rise
to further complexities and certain computational issues.

Other models also need to use approximate inferential methods such as
an Expectation Maximisation algorithm (EM) or Variational Bayes (VB)
which are still quite difficult to implement for these classes and add
other problems to the model selection process. They also often use a
bootstrap analysis to verify if the effective connectivity is
significant dramatically slowing down any search. Thus, searching over
these classes becomes more difficult and time consuming: a particular
problem is that here we are selecting from a large set of alternative
hypotheses. So LMDM provides a very promising fast and informative
explanatory data analysis of the nature of the dynamic network.

Another important advantage of the LMDM over most of its competitors is
that it comes with a customised suite of diagnostic methods. We
demonstrate some of these below.

\section{Scoring the MDM Using an Integer Programming Algorithm}
\label{IPA}

It is well known that finding the highest scoring model even within the
class of vanilla BNs is challenging. Even after using prior information
to limit this number to scientifically plausible ones, it is usually
necessary to use search algorithms to guide the selection. However
recently there have been significant advances in performing this task
\citep[see \emph{e.g.}][]{Meek97,spirtes2000causation,ramsey2010six,Cussens10,cowell2013simple},
and below we make use of one of the most
powerful methods currently available. We exploit the additive nature of
the MDM score function --- equation (\ref{mod}), where there are
exactly $n$ terms, one per region. Each region has $2^{n-1}$ possible
configurations, according to whether each other region is included or
excluded as a parent. Thus exhaustive computation of all possible score
{\em components} is feasible; for example, a 10 node network has only
5,120 possible components. Using the constrained integer programming
method described below, we have a method that allows the selection of
the optimal MDM with only modest computational effort. However,\vadjust{\eject} because
of the fully Bayesian formulation of the processes, it is also possible
to adapt established predictive diagnostics to this domain to further
examine the discrepancies associated with the fit of the best scoring
model and adjust the family where necessary. How this can be done is
explained in the following section, and the results of such an analysis
are illustrated in Section \ref{Real}.

Model selection algorithms for probabilistic graphical models can be
classified into two categories: the \emph{constraint-based method} and
the \emph{search-and-score method}. The former uses the conditional
independence constraints whilst the latter chooses a model structure
that provides the best trade-off between fit to data and model
complexity using a scoring metric. For instance, the PC (``Peter and
Clark") algorithm is a constraint-based method and searches for a
partially directed acyclic graph \citep[PDAG;][]{meek1995causal,spirtes2000causation,kalisch2008robustification}.
A PDAG is a graph
that may have both undirected and directed edges but no cycles. This
algorithm searches for a PDAG that represents a Markov equivalence
class, beginning with a complete undirected graph. Then, edges are
gradually deleted according to discovered conditional independence.
This means that edges are firstly deleted if they link variables that
are unconditionally independent. The same applies if the variables are
independent conditional on one other variable, conditional on two other
variables, and so on. In contrast, Greedy Equivalence Search (GES) is a
search-and-score method using the Bayes Information Criterion \citep
[BIC;][]{schwarz1978estimating} to score the candidate structures \citep
{Meek97,chickering2003optimal}. The algorithm starts with an empty
graph, in which all nodes are independent and then gradually, all
possible single-edges are compared and one is added each time. This
process stops when the BIC score no longer improves. At this point, a
reverse process is then driven in which edges are removed in the way
described above. Again, when the improvement of the score is not
possible, the graphical structure that represents a DAG equivalence
class is chosen \citep{ramsey2010six}. Note that as PC and GES search
for a Markov equivalence class, it is not possible to use them with
MDM, which discriminates graphical structures that belong to the same
equivalence class.

~~~\\
\noindent
\emph{The Integer Programming Algorithm}

The insight we use in this paper is that the problem of searching
graphical structures for MDM can be seen as an optimization problem
suitable for solving with an \emph{integer programming} (IP) algorithm.
Here, we use IP for the first time to search for the graphical
structure for the MDM, adapting established IP methods used for BN
learning. \citet{Cussens10} developed a search approach for the BN
pedigree reconstruction with the help of auxiliary integer-valued
variables, whilst \citet{cowell2013simple} used a dynamic programming
approach with a greedy search setting for this same problem. \citet
{jaakkola10} also applied IP, but instead of using auxiliary variables,
they worked with \emph{cluster-based constraints} as explained below.
\citet{cussens11} took a similar approach to \citet{jaakkola10} but
with different search algorithms. The MDM-IP algorithm follows the
approach provided by \citet{cussens11}, but with the MDM scores given
by LPL rather than BN scores, as we will show below.

Recall from Section \ref{mdm} that we use the joint log predictive
likelihood (LPL) to score candidate models and that this likelihood has
a closed form as a product of multivariate Student t distributions. By
equation \ref{mod}, the joint log predictive likelihood can be written
as the sum of the log predictive likelihood for each observation series
given its parents: a \emph{modularity} property \citep
{heckerman1998tutorial}. This assumption says that the predictive
likelihood of a particular node depends on only the graphical
structure, \emph{i.e.}, if the set of parents of node $i$ in $m_1$ is
the same as in $m_2$, then $LPL(\mathbf{Y}(i)|m_1)=LPL(\mathbf
{Y}(i)|m_2)$. Therefore, for any candidate
model $m$, $\mathrm{LPL}(m)$ is a sum of $n$ `local scores', one for
each node $r$, and the local score for $Y_{t}(r)$ is determined by the
choice of parent set $Pa_{m}(r)$ specified by the model $m$. Let
\localscore{r}{Pa_{m}(r)}{}=$\sum_{t=1}^T \log p_{tr}(y_t(r)|\mathbf
{y}^{t-1}, Pa_{m}(r))$, the local score, so that
$
\mathrm{LPL}(m) = \sum_{r=1}^{n} \localscore{r}{Pa_{m}(r)}
$.

Rather than viewing the model selection for the MDM directly as a
search for a model
$m$, we view it as a
search for $n$ subsets $\varsubset(1), \dots\varsubset(n)$ which
maximise $\sum_{r=1}^{n} \localscore{r}{\varsubset(r)}$ subject to
there existing an MDM model $m$ with $\varsubset(r) = Pa_{m}(r)$ for
$r=1,\dots n$.
We thus choose to see model selection as a problem of constrained
discrete optimisation. In the first step of our
approach we compute local scores $\localscore{r}{\varsubset}$ for all
possible values of $\varsubset$ and $r$. Next we create indicator variables
$\fv{r}{\varsubset}$, one for each local score. $\fv{r}{\varsubset}=1$
indicates that $Pa_{m}(r)=\varsubset$ in some candidate model $m$.
Note that creating all these local scores and variables is
practical considering the number of nodes in this application. The
model selection problem can now be posed in terms of the $\fv
{r}{\varsubset}$ variables:

\begin{quotation}
\normalsize
Choose values for the \fv{r}{\varsubset} variables to maximise
\begin{equation}
\label{eq:obj}
\sum\localscore{r}{\varsubset} \fv{r}{\varsubset}
\end{equation}
subject to there existing an MDM model $m$ with $\fv{r}{\varsubset}=1$
iff $\varsubset=Pa_{m}(r)$.
\end{quotation}

We choose an \emph{integer programming (IP)} \citep{Wolsey}
representation for this problem. To be an IP problem the objective
function must be linear, and all variables must take integer values.
Both of these are indeed the case in this application. However, in
addition, all constraints on solutions must be linear---an issue which
we now consider.

Clearly, any model $m$ determines exactly one parent set for each
$Y_{t}(r)$. This is represented by the following $n$ linear
\emph{convexity constraints}:
\begin{equation}
\label{eq:convexity}
\forall r = 1, \dots n: \sum_{\varsubset} \fv{r}{\varsubset} = 1.
\end{equation}
It is not difficult to see that constraints (\ref{eq:convexity}) alone
are enough to ensure that any solution to our IP problem represents a
directed graph (\emph{digraph}). Additional constraints are required
to ensure that any such graph is \emph{acylic}.

There are a number of ways of ruling out cyclic digraphs. We have
found the most efficient method is to use \emph{cluster constraints} first
introduced by \citet{jaakkola10}. These
constraints state that in an acylic digraph\vadjust{\eject} any subset (`cluster') of
vertices must have at least one member with no parents in that
subset. Formally:
\begin{equation}
\label{eq:cluster}
\forall\cluster\subseteq\{1, \dots, n\}:
\sum_{r \in\cluster}
\sum_{\varsubset: \varsubset\cap\cluster= \emptyset}
\fv{r}{\varsubset} \geq1.
\end{equation}

Maximising the linear function (\ref{eq:obj}) subject to linear
constraints (\ref{eq:convexity}) and (\ref{eq:cluster}) is an IP
problem. For values of $n$ which are not large, such as those
considered in the current paper, it is possible to explicitly represent
all linear constraints. An off-the-shelf IP solver such as CPLEX \citep
{cussens11} can then be used for model selection.

To solve our IP problem we have used the \gobnilp{} system \citep
{cussens11,bartlett13} which uses the \scip{} IP framework \citep
{Achterberg}. In \gobnilp{} the convexity constraints (\ref
{eq:convexity}) are present initially but not the cluster constraints
(\ref{eq:cluster}). As is typical in IP solving, \gobnilp{} first
solves the \emph{linear relaxation} of the IP where the \fv
{r}{\varsubset} variables are allowed to take any value in $[0,1]$ not
just 0 or 1. The linear relaxation can be solved very quickly. \gobnilp
{} then searches for cluster constraints which are violated by the
solution to the linear relaxation. Any such cluster constraints are
added to the IP (as so-called \emph{cutting planes}) and the linear
relaxation of this new IP is then solved and cutting planes for this
new linear relaxation are then sought, and so on. If at any point the
solution to the linear relaxation represents an acylic digraph, the
problem is solved. It may be the case that no cluster constraint
cutting planes can be found, even though the problem has not been
solved. In this case \gobnilp{} resorts to branch-and-bound search. In
all cases, we are able to solve the problem to optimality, returning an
MDM model which is guaranteed to have maximal joint log predictive
likelihood (LPL).

~~~\\
\noindent
\emph{The Running Time of the MDM-IPA}

The learning network process follows two steps. Initially, the scores
for each set of parents for individual nodes are found. Then the
MDM-IPA is applied to discover the best MDM over all nodes.

The run-time of the first step (finding the scores) of course depends
critically on the number of nodes and the sample size. It is necessary
to fit a linear dynamic model for every node and every set of parents
--- there are $2^{n-1}$ possible sets of parents per node. In addition,
when the observational variance is unknown, it is necessary to fit
every model several times, according to different values of the
discount factor (DF). Then the model (with a particular value of DF)
which provides the highest score is selected.

Table \ref{table_time} shows the time taken in minutes to find the
scores for different numbers of nodes and sample size, on a 2.7 GHz
quad-core Intel Core i7 linux host with 16 GB, using the software
\let\footnote\relax{R}\footnotemark[1]\footnotetext[1]{{\tt http://www.r-project.org/}}.
The discount factor was chosen in the range from 0.5 to 1.0 with
increments of 0.01. There is a sharp increase in the process time when
the underlying graph has 11 or more nodes.

The application of the IPA to scores found in\vadjust{\eject} the first step is usually
fast. For 11-node networks, the IPA took around 30 seconds using the
software \let\footnote\relax\gobnilp\footnotemark[2]\footnotetext[2]{{\tt http://www.cs.york.ac.uk/aig/sw/gobnilp/}}.

\begin{table}
\begin{center}
\begin{tabular}{c|cccc}
\hline
The number of & ~ & Sample Size ($T$)& ~ & ~ \\
nodes ($n$) & 100 & 200 & 600 & 1200 \\ \hline
3 & 0.13 & 0.25 & 0.75 & 1.51\\
4 & 0.35 & 0.68 & 2.05 & 4.09 \\
6 & 3.79 & 7.39 & 21.15 & 39.94 \\
11 & 167.99 & 325.50 & 1001.87 & 1982.08 \\ \hline
\end{tabular}
\caption{The time in minutes to find the scores for different numbers
of nodes and sample sizes, and the discount factor was chosen in the
range from 0.5 to 1.0 with increments of 0.01.}
\label{table_time}
\end{center}
\end{table}

~~~\\
\noindent
\emph{An Application of the MDM-IPA Using a Synthetic fMRI Dataset}

We now demonstrate the competitive performance of the MDM-IPA using a
simulated fMRI time series data \citep{smith2011network}. These data
were generated using the DCM fMRI forward model and considering some of
the characteristics that have been found in typical fMRI data analysed
so far, \emph{e.g.}\ set so that the amplitude of the neural time
series is of a typical magnitude. Note that because the simulation was
not driven by an MDM, we could not know a priori that this class of
model would not necessarily fit this dataset well. It therefore
provided a rigorous test of our methods within the suite of simulates available.

\begin{figure}[!htb]
\centering
\includegraphics{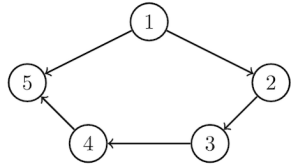}
\caption{The graphical structure used by \citet{smith2011network} to
simulate data.}
\label{fig3ni}
\end{figure}

Here we chose the dataset \emph{sim22} from \citet{smith2011network},
which has 5 regions, 10min-session, time resolution (\emph{i.e.} sample
rate) of 3.00s, 50 replications and the same graphical structure (see
Figure \ref{fig3ni}). The connection strength was defined according to
a random process and therefore varies over time, as \citeauthor
{smith2011network} explain: ``The strength of connection between any
two connected nodes is either unaffected, or reduced to zero, according
to the state of a random external bistable process that is unique for
that connection. The transition probabilities of this modulating input
are set such that the mean duration of interrupted connections is
around 30s, and the mean time of full connections is about 20s."

The log predictive likelihood (LPL) was first computed for different
values of discount factor $\delta$, using a weakly informative prior
with $n_0(r)=d_0(r)=0.001$ and $\mathbf{C}^*_0(r)=3\mathbf{I}_{p_r}$
for all $r$. The discount factors were chosen as the value that
maximised the LPL, and so the average DF over all replications and
nodes was around $0.85$ (smaller than $1$). Note that the MDM correctly
identified that connectivities have been simulated to vary over time.

%

\citet{smith2011network} compared different connectivity estimation
methods ranging from the simplest approach which only considered
pairwise relationships, such as correlation across the variables in the
different time series, to complex approaches\vadjust{\eject} which estimated a global
network using all nodes simultaneously, such as BNs. The main measures
that they used to compare these methods were \emph{c-sensitivity} and
\emph{d-accuracy}. The former represents the ability of the method to
correctly detect the presence of the connection, while the latter shows
the ability of methods to distinguish the directionality of the
relation between the nodes.

The first measure calculated was c-sensitivity as a function of the
estimated strength connectivity of the \emph{true positive} (TP) edges
which exist in both the true and the estimated graph, regardless of the
directionality, and the \emph{false positive} (FP) edges that exist in
the estimated graph but not in the true DAG. Here, we assess the
performance of the methods in detecting the presence of a network
connection, using the following measures:
\begin{itemize}
\item\emph{Sensitivity} $=\#TP/(\#TP+\#FN)$, where $\#$ represents
``the number of'' and FN is an abbreviation for false negative edge
which is a true connection that does not appear in the estimated graph.
This measure represents the proportion of true connections which are
correctly estimated;\\
\item\emph{Specificity} $=\#TN/(\#TN+\#FP)$, where TN is an
abbreviation for a true negative edge which does not exist in both true
and estimated graphs: \emph{i.e.} the proportion of connections which
are correctly estimated as nonexistent;\\
\item\emph{Positive Predictive Value} $=\#TP/(\#TP+\#FP)$: \emph{i.e.}
the proportion of estimated connections which are in fact true;\\
\item\emph{Negative Predictive Value} $=\#TN/(\#TN+\#FN)$: \emph{i.e.}
the proportion of connections estimated as nonexistent that do not
exist in the true graph;\\
\item\emph{Success Rate} $=(\#TP+\#TN)/10$, where $10$ is the total
number of possible connections for an undirected graph with 5 nodes.
This represents the proportion of correctly estimated connections.
\end{itemize}

\begin{figure}
\centering
\includegraphics{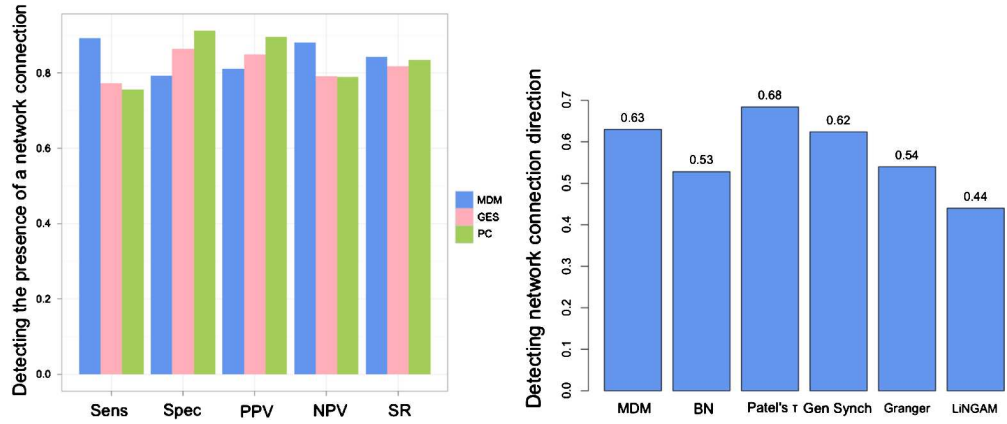}
\caption{\small{(\emph{left}) The average over 50 replications of the
\emph{sensitivity} (Sens) $=TP/(TP+FN)$; \emph{specificity} (Spec)
$=TN/(TN+FP)$; \emph{positive predictive value} (PPV) $=TP/(TP+FP)$;
\emph{negative predictive value} (NPV) $=TN/(TN+FN)$; (SR) \emph
{success rate} $=(TP+TN)/(\text{total number of connections})$ for
three methods: MDM (blue bar), GES (salmon bar) and PC (green bar).
(\emph{right}) The average over 50 replications of the percentage of
directed connections that was detected correctly for some methods. The
results of this second figure are from \citet{smith2011network}, except
for the method MDM.}}\vspace*{-6pt}
\label{fig4s22} 
\end{figure}

According to \citet{smith2011network}, on the basis of c-sensitivity,
the best methods are algorithms that use the Bayesian Network models.
We therefore implemented two methods: the GES and the PC in the Tetrad \let\footnote\relax
IV\footnotemark[1]\footnotetext[1]{{\tt http://www.phil.cmu.edu/projects/tetrad/current.html}} (see the
definition of these methods in the beginning of this section).\ The
implementation of these methods is fairly easy, but unsurprisingly the
computational time of the MDM is considerably higher than others
because its descriptive search space is much larger. We estimated our
sensitivity measures, as described above, and Figure \ref{fig4s22}
(left) shows the average sensitivity measures over $50$ replications
for the MDM-IPA (blue bar), the GES (salmon bar) and the PC (green bar)
methods. These approaches show satisfactory results for all measures,
with the mean percentage above $75\%$. Although the PC has the highest
percentage in Specificity and Positive Predictive Value, the MDM
performs better in the three other measures. For instance, the MDM
correctly detected around $90\%$ of the true connections whilst PC and
GES detected about $75\%$ (sensitivity measure). Moreover, the MDM has
the highest overall percentage of correct connections (success rate).

As a second method of comparison, \citet{smith2011network} proposed a
way to compare the performance of the methods in detecting the \emph
{direction} of connectivity. The d-accuracy is calculated as the
percentage of directed edges that are detected correctly. This measure
is given in Figure \ref{fig4s22} (right). Again the MDM obtained some
of the best results for this measure. Other methods that also had good
results according to this criterion were Patel's measures \citep
{patel2006bayesian} and Generalised synchronization \citep[Gen
Synch;][]{quiroga2002performance}. We note that the performance of
LiNGAM was poor when compared with other methods.

Although the d-accuracy of Patel's $\tau$ and Gen Synch is not
substantially different from that of the MDM (Figure \ref{fig4s22},
right), these two former methods have only moderate c-sensitivity
scores \citep{smith2011network}. The opposite pattern can be seen for
the methods based on the BN: they perform well in c-sensitivity but
poorly in d-accuracy. Thus, only the MDM performed well in all the
measures at the individual level of analysis.

~~~\\
\noindent
\emph{An MDM Synthetic Study}

We have showed the performance of the MDM-IPA considering synthetic
data from 5-node networks. It would be interesting to see how this
search algorithm performs with a larger number of nodes. However,
although \citet{smith2011network} provided higher dimensional DCM fMRI
synthetic data, none of these were generated considering the stochastic
process in connectivity strengths. We therefore generated the MDM data
based on our analysis of the resting-state experiment studied in
Section \ref{Real}. More specifi-cally, we fixed the $11$ nodes and
$230$ time points in this experiment and simulated data of this size
assuming as true the best fitting MDM model we found for the original
data set (green edges in Figure \ref{Figsubj7}). More details about the
simulation process are described in Appendix A.\vadjust{\vfill{\eject}}

\begin{figure}[h!]
\centering
\includegraphics{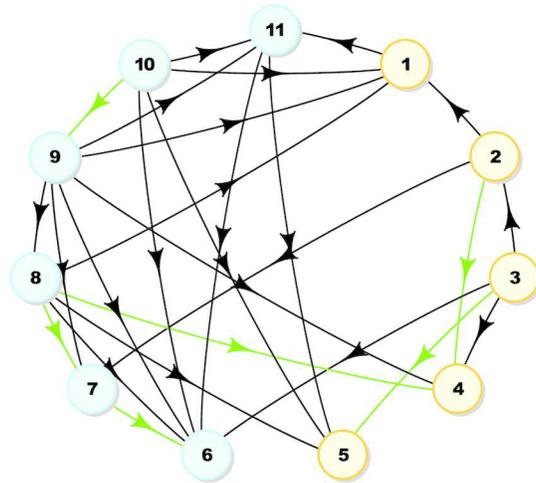}
\caption{\small{The graphical structure estimated for subject $7$ using
the MDM-IPA. The green edges are the significant connectivities found
in the group analysis (see Figure \ref{ba_fig1a} (b) in Section 6 below).}}
\label{Figsubj7}\vspace*{-6pt}
\end{figure}

Algorithms GES, PC and MDM-IPA were applied for 50 replications. Using
a weakly informative prior for the MDM and considering the network
estimated by the MDM-IPA, the average DF over replications was $0.83$.
This is very close to the one found in 5-node networks study\vadjust{\eject} ($0.85$).
In this sense, both sets of synthetic data (the DCM with 5 nodes and
the MDM with 11 nodes) have a similar variability of connections over
time. When \citet{smith2011network} compared the results of data over
different numbers of nodes, they concluded that the classification
order of methods considering c-sensitivity and d-accuracy measures is
extremely similar for 5, 10, and 15 nodes. Here, considering the
algorithms GES, PC and MDM-IPA, we came to the same conclusion. Table
\ref{table_11nod} shows that in general the MDM-IPA has the highest
c-sensitivity measures and much better scores. While almost $80\%$ of
the estimated connections are actually true (PPV measure) for the
MDM-IPA in both sets of synthetic data, for the GES this percentage was
$85\%$ in 5-node networks and it decreased to around $20\%$ in 11-node
networks (the PC provided a similar pattern). A plausible reason for
this is that the number of false positive connections is dramatically
higher for the GES and the PC than for the MDM-IPA in 11-node networks,
\emph{i.e.} the average $\#FP$ over replications for the GES, the PC
and the MDM-IPA, respectively, was around $0.7$, $0.4$ and $1.0$ in
5-node networks, and around $10$, $18$ and $2$ in 11-node networks.
Although the prevalence of FP increases with the number of nodes, their
connectivity strengths usually are close to zero, as shown below.

\begin{table}[htbp]
\centering
\begin{tabular}{cccc}
\toprule
c-sensitivity & MDM-IPA & GES & PC \\
\midrule
Sens & 0.83 & 0.85 & 0.74 \\
Spec & 0.97 & 0.64 & 0.80 \\
PPV & 0.82 & 0.23 & 0.32 \\
NPV & 0.98 & 0.97 & 0.96 \\
SR & 0.96 & 0.67 & 0.79 \\
\bottomrule
\end{tabular}
\caption{\small{The average over 50 replications of the \emph
{sensitivity} (Sens) $=TP/(TP+FN)$; \emph{specificity} (Spec)
$=TN/(TN+FP)$; \emph{positive predictive value} (PPV) $=TP/(TP+FP)$;
\emph{negative predictive value} (NPV) $=TN/(TN+FN)$; (SR) \emph
{success rate} $=(TP+TN)/(\text{total number of connections})$ for
three methods: the MDM-IPA, the GES and the PC, considering 11-node
networks synthetic data.}}
\label{table_11nod}
\end{table}

Regarding the d-accuracy criteria, the MDM-IPA also demonstrated
greater power in detecting the direction of connectivity than the GES
and the PC --- $60\%$, $45\%$ and $26\%$ of directed edges were
detected correctly in 11-node networks for the MDM-IPA, the GES and the
PC, respectively.

In addition, we evaluated the proportion of time $PT_{ri}$ that the
true value of connection $i$ for node $r$ is inside the $95\%$ smoothed
highest posterior density (HPD) interval. As a result, the average of
$PT_{ri}$ over all replications, considering only TP connections (green
edges in Figure \ref{Figsubj7}), was $96\%$, and therefore the MDM-IPA
was shown to be efficient not only in detecting the edges, but also in
estimating the connectivity strengths.

There are two kinds of FP connections: (\emph{situation 1}) the edges
exist in both the true and the estimated networks, but with opposite
directions (\emph{e.g.} connection $9 \rightarrow10$ exists in the
estimated graph whilst the true connection is $10 \rightarrow9$), and
(\emph{situation 2}) the edges exist only in the estimated network
(\emph{e.g.} connection $4 \rightarrow10$). Considering the true value
of these FP regression parameters as zero, the average $PT_{ri}$ over
all replications, considering the FP connections with opposite
directions in true network (\emph{situation 1}) was about $30\%$. This
is no surprise, when the opposite connection strength is higher than
zero. In contrast, the average $PT_{ri}$ over all replications,
considering the FP connections that do not exist even on an undirected
true network (\emph{situation 2}) was $75\%$. This shows that when the
MDM-IPA provides spurious connections, the associated connectivity
strengths are usually close to zero. It is interesting to note that
this appearance of spurious but weak dependences is a well known
phenomenon when fitting more standard graphical models using Bayes
factor methods. More robust conjugate Bayes model selection methods
have recently been investigated using non-local priors \citep[see \emph
{e.g.}][]{Consonni}, and analyses of these could provide promising
alternatives to the scoring methods in this paper.

\section{The Use of Diagnostics in an MDM}
\label{diag}

In the past, \citet{Cowell99} have convincingly argued that when
fitting graphical models, it is extremely important to customise
diagnostic methods, not only to determine whether the model appears to
be capturing the data generating mechanism well but also to suggest
embellishments of the class that might fit better. Their preferred
methods are based on a one step ahead prediction. We use these here.
They give us a toolkit of sample methods for checking to see whether
the best fitting model we have chosen through our selection methods is
indeed broadly consistent with the data we have observed. We modified
their statistics to give analogous diagnostics for use in our dynamic
context. We give three types of diagnostic monitor, based on analogues
for probabilistic networks \citep{Cowell99}.

First the \emph{global monitor} is used to compare networks. After
identifying a DAG providing the best explanation over the LMDM
candidate models, the predicted relationship between a particular node
and its parents can be explored through the \emph{parent-child
monitor}. Finally the \emph{node monitor} diagnostic can indicate
whether the selected model fits adequately. If this is not so, then a
more complex model will be substituted, as illustrated below.

~~~\\
\noindent
\emph{I - Global Monitor}

The first stage of our analysis is to select the best candidate DAG
using simple LMDMs, as described in Section \ref{IPA}. It is well known
that the prior distributions on the hyperparameters of candidate models
sharing the same features must first be matched \citep
{heckerman1998tutorial}. In this way, the BF techniques can be
successfully applied in the selection of non-stochastic graphs in real
data. If this is not done, then one model can be preferred to another,
not for structural reasons but for spurious ones. This is also true for
the dynamic class of models we fit here.

However, fortunately, the dynamic nature of the class of the MDM
actually helps dilute the misleading effect of any such mismatch
because after a few time steps, evidence about the conditional
variances and the predictive means is discounted and the marginal
likelihood of each model usually repositions itself. In particular, the
different priors usually have only a small effect on the relative
values of subsequent conditional marginal likelihoods. We describe
below how we have nevertheless matched priors to minimise this small
effect in the consequent Bayes factor scores driving the model selection.

Just as for the BN to match priors, we can exploit a decomposition of
the Bayes factor score for the MDMs. Because of the modularity
property, when some features are incorporated within the model class,
the \emph{relative} score of such models only discriminates the
components of the model where they differ. Thus, consider again the
graphical structure in Figure \ref{fig5new}. For instance, suppose the
LMDM is updated because node $3$ exhibits heteroscedasticity. On
observing this violation, the conditional one-step forecast
distribution for node $3$ can be replaced by one relating to a more
complex model. Thus, a new one step ahead forecast density,
$p^*_{t3}(y_t(3)|\mathbf{y}^{t-1}, y_t(1), y_t(2))$, is $(Y_t(3) |
\mathbf{y}^{t-1}, y_t(1), y_t(2)) \sim\mathcal{T}_{n_{t-1}(3)}(f_t(3),
Q^h_t(3))$, where the parameters $f_t(3)$ and $n_{t-1}(3)$ are defined
as before, but $Q^h_t(3)$ is now defined as a function of random
variance, say $k_{t3}(\mathbf{F}_t(3)' \boldsymbol{\theta}_t(3))$ \citep
[see details in ][section 10.7]{West}. The log Bayes factor comparing
the original model with a heteroscedastic model is calculated as
\begin{eqnarray*}
\log\text{(BF)} = \sum_{t=1}^T \log p_{t3}(y_t(3)|\mathbf{y}^{t-1},
y_t(1), y_t(2)) - \sum_{t=1}^T \log p_{t3}^*(y_t(3)|\mathbf{y}^{t-1},
y_t(1), y_t(2)).
\end{eqnarray*}

We set prior densities over the same component parameters over
different models, because the model structure is common for all other
nodes. The BF then discriminates between two models by finding the one
that best fits the data \emph{only} from the component where they
differ: in our example the component associated with node $3$. Even in
larger scale models like the ones we illustrate below, we can therefore
make a simple modification of scores in order to use the IP algorithm
derived above, and in this way adapt the scores over graphs almost
instantaneously.

In this setting we have found that the distributions for
hyperparameters of different candidate parent sets is not critical for
the BF model selection, provided that early predictive densities are
comparable. We have found that a very simple way of achieving this is
to set the prior covariance matrices over the regression parameters of
each model a priori so that they are independent with a shared
variance. Note the hyperparameters and the parameter $\delta$ of the
nodes 1, 2 and 4 were the same for both models: homoscedastic and
heteroscedastic for node 3. Many numerical checks have convinced us
that the results of the model selection we describe above are
insensitive to these settings \emph{provided} that the high scoring
models pass various diagnostic tests some of which we discuss below.

~~~\\
\noindent
\emph{An Application of the Global Monitor Using Synthetic Data}

Now we show an application of the global monitor in a simple example.
As we argued in Section \ref{mdm}, BNs with the same skeleton are often
Markov equivalent \citep{lauritzen1996graphical}. This is not so for
their dynamic MDM analogues. As a result it is possible for an MDM to
detect the directions of relationships in DAGs which are Markov
equivalent in static analysis. We will explore these issues below, and
demonstrate how this is possible using a simulation experiment. This in
turn allows us to search over hypotheses expressing the potential
deviations of causations.

\begin{figure}[!htb]
\centering
\includegraphics{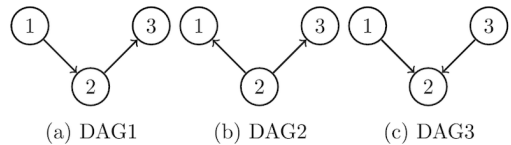}
\caption{Directed acyclic graphs used in the MDM synthetic study. (a)
DAG1 and (b) DAG2 are considered Markov equivalent whilst neither are
equivalent to (c) DAG3.}
\label{fig5ni}
\end{figure}

Here simulation of observations from known MDMs were studied using the
graphical structure DAG1 (Figure \ref{fig5ni} (a)), sample sizes
$T=100$, $200$ and $300$, and different dynamic levels $\mathbf{W}^*(r)
=0.001\mathbf{I}_{p_r}$ and $0.01\mathbf{I}_{p_r}$. The impact of these
different scenarios on the MDM results was verified regarding $3$
regions and $100$ datasets for each $T$ and $\mathbf{W}^*(r)$ pair.
Details about the simulation process can be seen in Appendix B.

\begin{figure}[h!]
\centering
\includegraphics{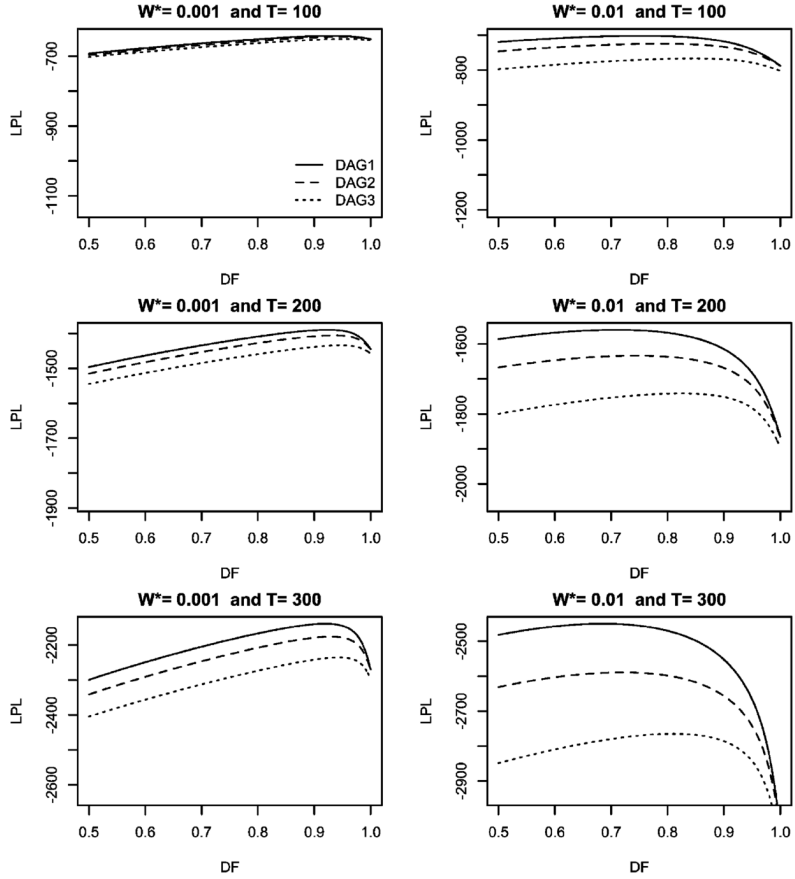}
\caption{\small{The log predictive likelihood versus different values
of discount factor (DF), for DAG1 (solid lines), DAG2 (dashed lines)
and DAG3 (dotted lines). The sample size increases from the first to
the last row whilst the dynamic level (innovation variance) increases
from the first to the last column. The range of the $y-$axis (LPL) is
the same size: $500$ for all graphs.}}
\label{fig8ni}\vspace*{-6pt}
\end{figure}

Figure \ref{fig8ni} shows the log predictive likelihood versus
different values of discount factor, considering DAG1 (solid lines),
DAG2 (dashed lines) and DAG3 (dotted lines). The sample size increases
from the first to the last row whilst the dynamic level (innovation
variance) increases from the first to the last column. Although the
ranges of the LPL differ across the graphs, the range sizes are the
same, \emph{i.e.} $500$ so that it is easy to compare them.

An interesting result is that when data follow a dynamic system but are
fitted by a static model, the non-Markov equivalent DAGs are
distinguishable whilst equivalent DAGs are not. For instance, when
$\mathbf{W}^*(r) = 0.01\mathbf{I}_{p_r}$ and $T=100$ (first row and
second column), the value of the LPL for DAG3 is smaller than the value
for the other DAGs, but there is no significant difference between the
values of the LPL for DAG1 and DAG2 when $\delta=1$, which we could
deduce anyway since these models are Markov equivalent \citep[see \emph
{e.g.}][]{Ali}. In contrast, there are important differences between
the LPL of DAGs when dynamic data are fitted with dynamic models
($\delta<1$), DAG1 having the largest LPL value. In particular, MDMs
appear to select the appropriate direction of connectivity with a high
success rate. However, their performance varies as a function of the
innovation variance and sample size (note how the distance between the
lines changes from one graph to another). For instance, as might be
expected, the higher the sample size, the higher the chance of
identifying the true DAG correctly. On the other hand, $T$ has the
greatest impact on the results when the dynamics of the data are very
slowly changing ($\mathbf{W}^*(r) = 0.001\mathbf{I}_{p_r}$). In this
case the percentage of replications in which the correct DAG was
selected was $40\%$, $80\%$ and $95\%$, for a sample size equal to
$100$, $200$ and $300$, respectively, whilst almost all the
replications selected the DAG correctly for whichever $T$ and $\mathbf
{W}^*(r) = 0.01\mathbf{I}_{p_r}$.

%

~~~\\
\noindent
\emph{II - Parent-child Monitor}

Because of the modularity property, the relationship between a
particular node and its parents can be assessed considering only this
component in the MDM. Let $Pa(r) = \{\mathbf{Y}_{pa(r)}(1), \dots,
\mathbf{Y}_{pa(r)}(p_r-1)\}$, then
\begin{equation*}
\log(BF)_{ri} = \log p_r\{\mathbf{y}(r)| Pa(r)\} - \log p_{r_i}\{\mathbf
{y}(r)| Pa(r) \setminus\mathbf{y}_{pa(r)}(i)\},
\end{equation*}
for $r=1, \dots, n$ and $i=1, \dots, p_r-1$; where $\{Pa(r) \setminus
\mathbf{Y}_{pa(r)}(i)\}$ means the set of all parents of $\mathbf
{Y}(r)$ excluding the parent $\mathbf{Y}_{pa(r)}(i)$.

\noindent
\emph{III - Node Monitor}

Again the modularity ensures that the model for any given node can be
embellished based on residual analysis. For instance, consider a
non-linear structure for a founder node $r$, \emph{i.e.} one with no
parents. On the basis of the partial autocorrelation of the residuals
of the logarithm of the series, a more sophisticated model of the form
\begin{eqnarray*}
\log Y_t(r) = \theta_t^{(1)}(r) + \theta_t^{(2)}(r) \log Y_{t-1}(r) + v_t(r),
\end{eqnarray*}
suggests itself. Note that this model still provides a closed form
score for these components. The lower scores and their corresponding
model estimation can then be substituted for the original steady models
to provide a much better scoring dynamic model, but they still respect
the same causal structure as in the original analysis.

Denoting $\log Y_t(r)$ by $Z_t(r)$, the conditional one-step forecast
distribution for $Z_t(r)$ can then be calculated using a DLM on the
transformed series $\{Z_t\}$. More generally if we hypothesize that
$Z_t(r)$ can be written as a continuous and monotonic function of
$Y_t(r)$, say $g(.)$, and so the conditional one-step forecast
cumulative distribution for $Y_t(r)$ can be found through
\begin{eqnarray*}
F_{Y_t(r)}(y) &=& P_{tr}(Y_t(r) \leq y|\mathbf{y}^{t-1},Pa(r))\\
&=& P_{tr*}(Z_t(r) \leq g^{-1}(y) |\mathbf{y}^{t-1},Pa(r))\\
&=& F_{Z_t(r)}(g^{-1}(y)).
\end{eqnarray*}

Thus $p^*_{tr}(y_t(r)|\mathbf{y}^{t-1},Pa(r))$, the conditional
one-step forecast density for $Y_t(r)$ for this new model can be
calculated explicitly \citep[see details in][section 10.6]{West}.

Recall that when the variance is unknown, the \emph{conditional}
forecast distribution is a noncentral $t$ distribution with a location
parameter, say $f_t(r)$, scale parameter, $Q_t(r)$, and degrees of
freedom, $n_{t-1}(r)$. The one-step forecast errors are defined as $e_t
(r)= Y_t(r) - f_t(r)$ and the standardised conditional one-step
forecast errors as $e_t (r)/Q_t(r)^{(1/2)}$. The assumption underlying
the DLM is that the standardised conditional one-step forecast errors
have an approximate Gaussian distribution, when $n_{t-1}(r)$ is large,
and they are serially independent with constant variance \citep{West,durbin2012time}.
These assumptions can be checked by looking at some
graphs, such as a QQ-plot, standardised residuals versus time,
cumulative standardised residuals versus time and the autocorrelation
function (ACF) plot \citep{smith1985diagnostic,harrison1991dynamic,durbin2012time,anacleto}.

\section{The Analysis of Real Resting-state fMRI Data}
\label{Real}

Finally, we demonstrate how our methods can be applied to a recent
experiment where the appropriate MDM needs more nodes than in our
previous examples so that the IPA becomes essential. We applied the
MDM-IPA network learning procedure to a resting-state fMRI dataset
\citep[described in detail in][]{duff2013utility}. Data were acquired
on $15$ subjects, and each acquisition consists of 230 time points,
sampled every 1.3 seconds, with 2x2x2 mm$^3$ voxels. The FSL \let\footnote\relax
software\footnotemark[2]\footnotetext[2]{{\tt
http://fsl.fmrib.ox.ac.uk}} was used for preprocessing, including head
motion correction, an automated artifact removal procedure \citep
{salimi2014automatic} and intersubject registration. We use $11$ brain
regions defined on 5 \emph{motor} and 6 \emph{visual} regions. Motor
nodes were selected based on activation patterns during a
finger-tapping task. The nodes were located within the cerebellum,
Putamen, Supplementary Motor Area (SMA), Precentral Gyrus and
Postcentral Gyrus (nodes numbered from 1 to 5 respectively). The visual
nodes were selected based on activation patterns during presentation of
abstract shapes in motion in the central visual field. The visual nodes
used are Visual Cortex V1, V2, V3, V4, V5 and task negative (regions in
peripheral V1 found to decrease with task in a separate study; nodes
numbered from 6 to 11 respectively). The observed time series are
computed as the average of fMRI data over the voxels of each of these
defined brain areas. We note that these data are not the output of an
ICA, which may confound the interpretation of the results.

\begin{figure}[b!]
\centering\vspace*{-12pt}
\includegraphics[scale=0.99]{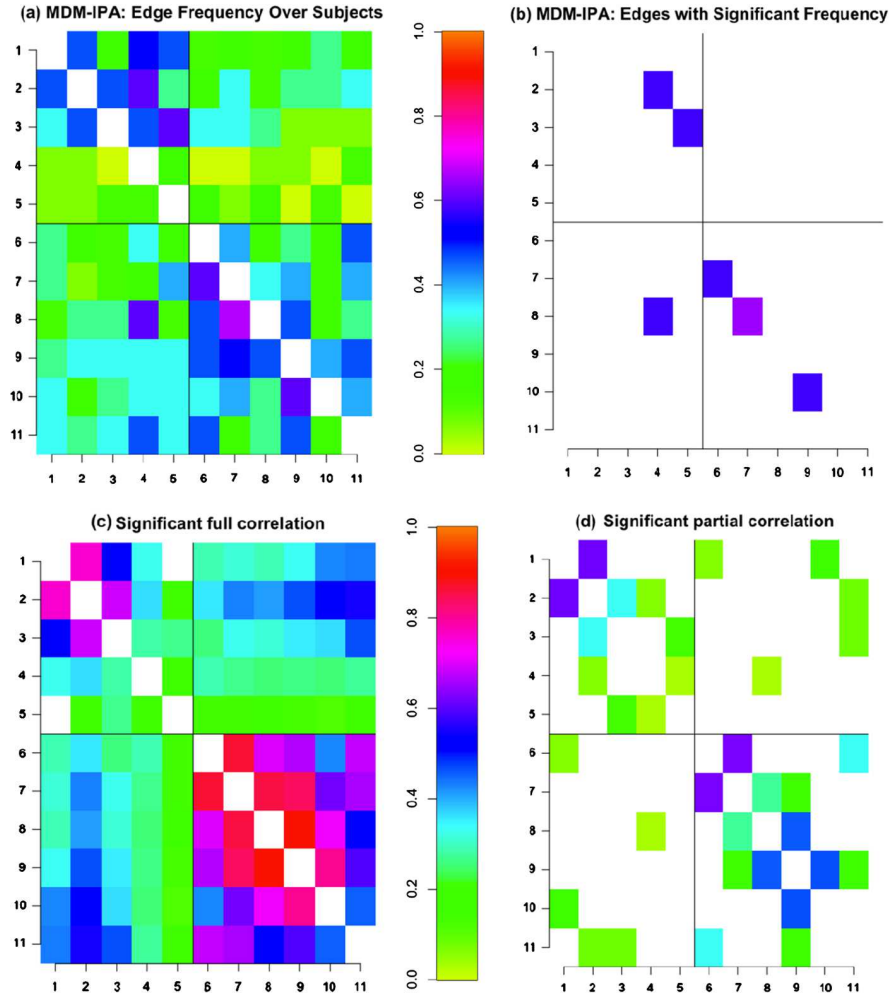}\vspace*{-6pt}
\caption{\small{The top row shows the proportion of subjects who have a
particular edge $i \rightarrow j$ (i-th row and j-th column) using (a)
the \emph{MDM-IPA} for all connectivities and (b) only for significant
connectivities. The bottom row shows the average of significant
correlation between two nodes across subjects using (c) \emph{full
correlation method} and (d) \emph{partial correlation}. Nodes 1-5 are
motor regions, while nodes 6-11 are visual regions; as expected, the
{\em intra-modal} connections (the 2 blocks on the diagonal) are more
prevalent and stronger than {\em cross-modal} connections. Within each
group, nodes are arranged according to the anticipated flow of
information in the brain.}}
\label{ba_fig1a}
\end{figure}

Here we modelled each subject using the MDM-IPA method, and then
compared the graphs of the brain connectivities across individuals.
Using a weakly informative prior, the scores of all possible sets of
parents for every node were found. The MDM-IPA was then used to
discover the optimal graphical structure to explain the data from each
subject. We assessed the intersubject consistency of the resulting
networks by the prevalence of directed edges and by verifying
completely homogeneous connectivity over the network. More
specifically, we estimated $p_{ij}$, the probability that an edge $i
\rightarrow j$ exists, as the proportion $\hat{p}_{ij}$ of subjects
with this particular edge among the identified regions. The statistic
we used measured the extent to which $p_{ij}>\pi$, where $\pi$ is the
edge occurrence rate under homogeneity, set equal to the average of
$\hat{p}_{ij}$ over the 90 possible edges. Figure \ref{ba_fig1a} (a)
shows $\hat{p}_{ij}$ for all connectivities $i \rightarrow j$, where
$i$ indexes rows and $j$ columns. Figure \ref{ba_fig1a} (b) also shows
$\hat{p}_{ij}$, but only for those edges significant with a $5\%$ false
discovery rate correction \citep[FDR;][]{benjamini1995controlling}. The
black horizontal and vertical lines divide the figure into four
squares; the top left square represents the connectivity between motor
brain regions, whilst the lower right square represents the one between
the visual brain regions. Unsurprisingly, most of connectivities are
within these two squares. The two other squares represent {\em
cross-modal} connections, between motor and visual regions which are
less prevalent.

We also consider two other methods of estimating the functional
connectivity: \emph{full correlation} and \emph{partial correlation}
\citep{baba2004partial,marrelec2006partial}. For each subject, for
each node pair, we computed the full and partial correlation, and thus
Figure \ref{ba_fig1a} (c) and (d) show the edges which appeared by
chance with a probability higher than $0.95$, respectively. Note that
these techniques provide symmetric results about the principal
diagonal. The vast majority of connections exist with high significance
(Figure \ref{ba_fig1a} (c)), however, connections with the strongest
correlation (above $0.6$) tend to be intra-modal as discussed above. As
expected, the significant MDM edges are a subset of the significant
partial correlations (Figure \ref{ba_fig1a} (d)). The nodes are ordered
according to the expected flow of information in the brain, and thus it
is notable that we find significant edges between consecutive nodes. In
short, while full and partial correlations do not account for
nonstationarities nor represent a particular joint model, Figure \ref
{ba_fig1a} demonstrates that the application of the MDM gives
scientifically plausible results.

%

%
\begin{figure}[t!]
\centering
\includegraphics{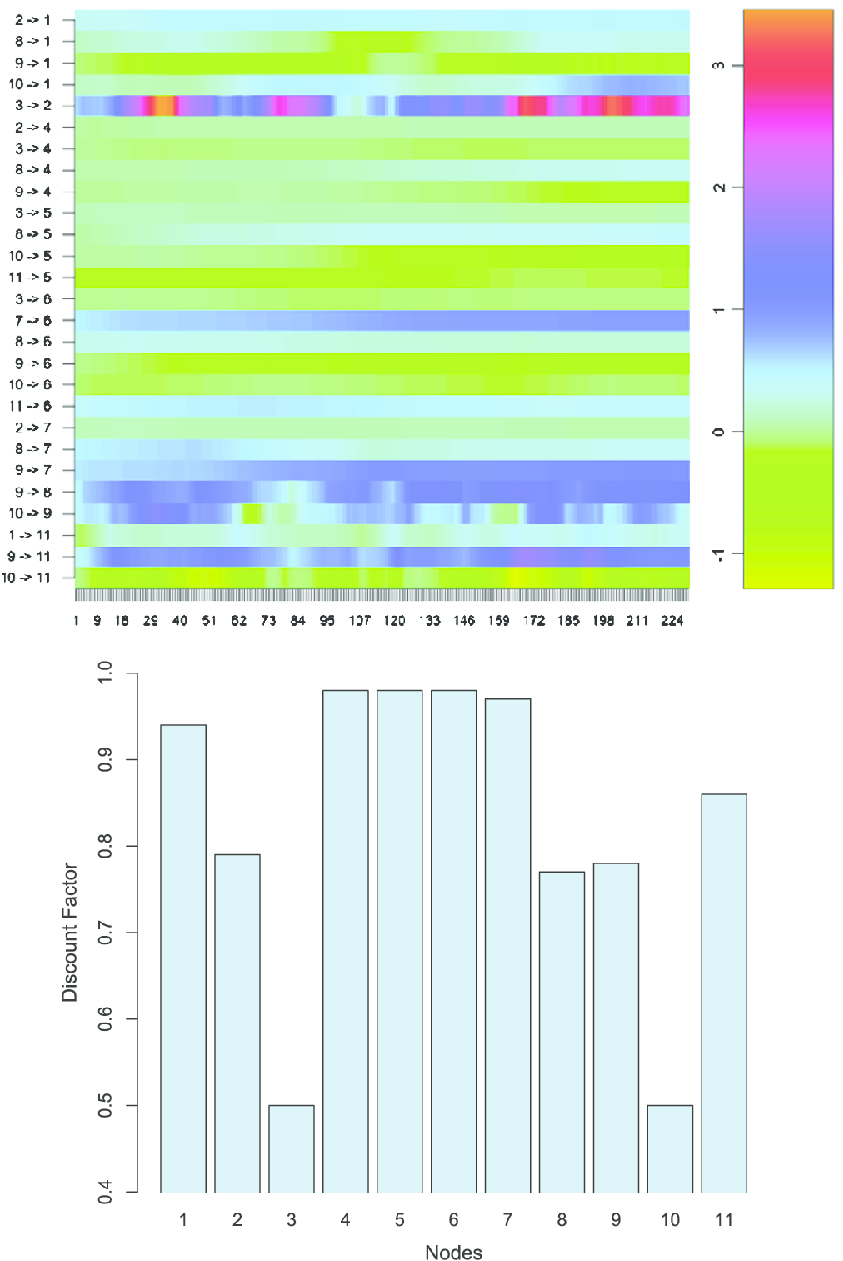}\vspace*{-6pt}
\caption{\small{\emph{(Above)} The smoothing posterior mean of
connectivities (in \emph{y-}axis) over time (in \emph{x-}axis). \emph
{(Bottom)} Discount factor for each node.}}
\label{ba_fig2}\vspace*{-6pt}
\end{figure}

To illustrate the use of the parent-child monitor and node monitor, we
selected subject $7$ because its MDM-IPA result contains all edges that
are significant in group analysis (Figure \ref{Figsubj7}), and so in
this sense it was a typical experimental subject. Figure~\ref{ba_fig2}
(above) provides the smoothed posterior mean for all connectivities
that exist in the graph of subject $7$ over time whilst the right
figure shows the discount factor found for every node. Note the
considerable variation in the connection strengths, for example from
region 3 to 2 (fifth connectivity shown in Figure \ref{ba_fig2},
above). This is consistent with other reports on the nonstationarities
of resting-state fMRI \citep{Allen,ge2009novel,leonardi2013principal}
and demonstrates a key capability of this model. One possible
explanation for the observed apparent changes in connectivity strengths
proposed by \citet{chang2010time} is that the level of attention,
arousal and daydreaming can differ during the resting-state experiment,
and this is reflected through the measurements.

The group network (Figure \ref{ba_fig1a} (b)) was fitted for all
subjects and as a result the patterns shown in Figure \ref{ba_fig2} are
consistent across subjects. For instance, the average DF for nodes that
have parents was $0.96$ for motor nodes and was $0.80$ for visual
nodes. It therefore appears that visual nodes have a shorter memory
than motor nodes. A possible reason is that the physical/sensory
environment is much more constrained/static than the visual
environment. With this experiment, subjects were shown a screen with a
fixation point. However, they were not explicitly asked to fixate. This
might explain the greater perceptual variability in visual relative to
sensory-motor areas.

\begin{figure}[t]
\centering
\includegraphics{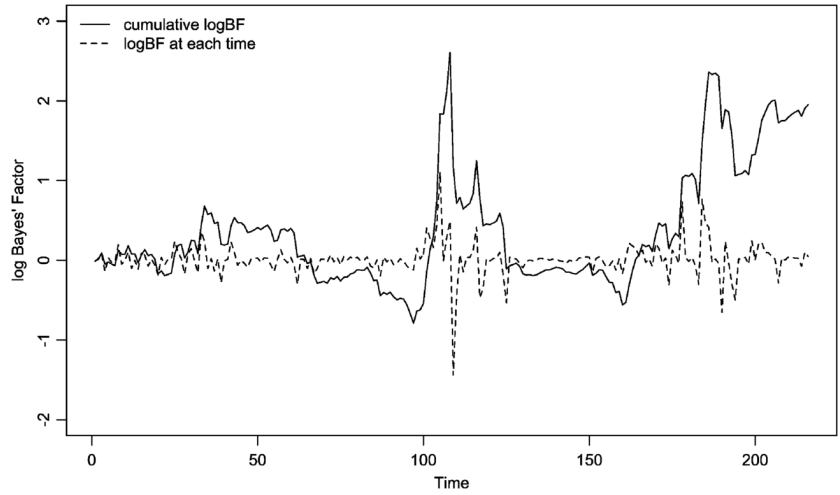}
\caption{\small{The logBF at each time (dashed lines) and cumulative
logBF (solid lines) comparing regions 2, 8, 9 and 10 as the set of
parents of region 1 with the same set of parents but without region 8.
The final logBF was $1.95$ and therefore, there is evidence for the
former model. This illustrates the use of the parent-child monitor.}}
\label{ba_fig5}
\end{figure}

We can diagnose and confirm the ``parent-child" relationships for
region $1$, as the connectivity from region $8$ into $1$ appears to be
near zero part of the time. The significance of this connectivity is
reflected in\vadjust{\eject}
\begin{eqnarray*}
\log(BF)_{12} &=& \log p_1\{\mathbf{y}(1)| \mathbf{y}(2), \mathbf{y}(8)
, \mathbf{y}(9) , \mathbf{y}(10)\} - \log p_{1_2}\{\mathbf{y}(1) |
\mathbf{y}(2) , \mathbf{y}(9) , \mathbf{y}(10)\},
\end{eqnarray*}
which, as discussed above (Section \ref{diag}), can be plotted over
time. Figure \ref{ba_fig5} shows the individual contributions to the
log(BF) as well as the cumulative log(BF). While the cumulative log(BF)
is sometimes close to zero, there is a surge in evidence for the larger
model (that includes 8 as a parent) near time point 110 and again after 180.

\begin{figure}[t]
\centering
\includegraphics{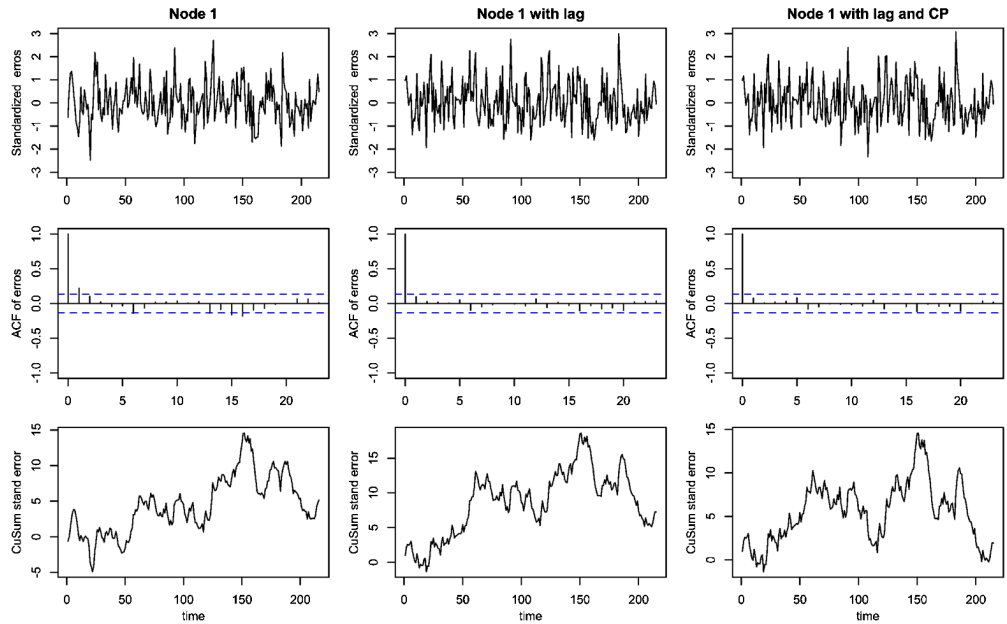}
\caption{\small{Time series plot, ACF-plot and cumulative sum of
one-step-ahead conditional forecast errors for region $1$ (first
column), considering \emph{lag 1} (second column) and considering \emph
{lag 1} and \emph{change points} (third column). This illustrates the
use of the node monitor.}}
\label{ba_fig3}
\end{figure}

As discussed above, a simple LMDM can easily be embellished in order to
solve problems detected by diagnostics measures. For example, Figure
\ref{ba_fig3} (first column) shows the time series, ACF and cumulative
sum plot of the standardised conditional one-step forecast errors for
node $1$. Note that the ACF-plot suggests autocorrelation at lag 1.
This feature can still be modelled within the MDM class by making a
local modification. For example, the past of region $1$ may be included
in its observation equation. That is,
\begin{eqnarray*}
Y_t(1) &=& \theta_t^{(1)}(1) + \theta_t^{(2)}(1)Y_t(2) + \theta
_t^{(3)}(1)Y_t(8) + \theta_t^{(4)}(1)Y_t(9) + \theta_t^{(5)}(1)Y_t(10)
\\
&~& + \theta_t^{(6)}(1)Y_{t-1}(1) + v_t(1).
\end{eqnarray*}

Figure \ref{ba_fig3} (second column) provides the residual analysis
plots considering the model with the lag 1. We can see that the
insertion of the past of the observation variable improves the
ACF-plot. However, the cumulative sum of forecast errors (second column
and third row) exhibits a non-random pattern, which suggests an
additional feature: the presence of change points. \citet[ch.\
11]{West} suggested a simple method to model this phenomenon as
follows. Firstly, the BF or the cumulative BF is calculated in each
time point comparing two models. If this measure is less than a
particular threshold, a new model is fitted which entertains the
possibility that a change point may have occurred.

\begin{figure}[t]
\centering
\includegraphics{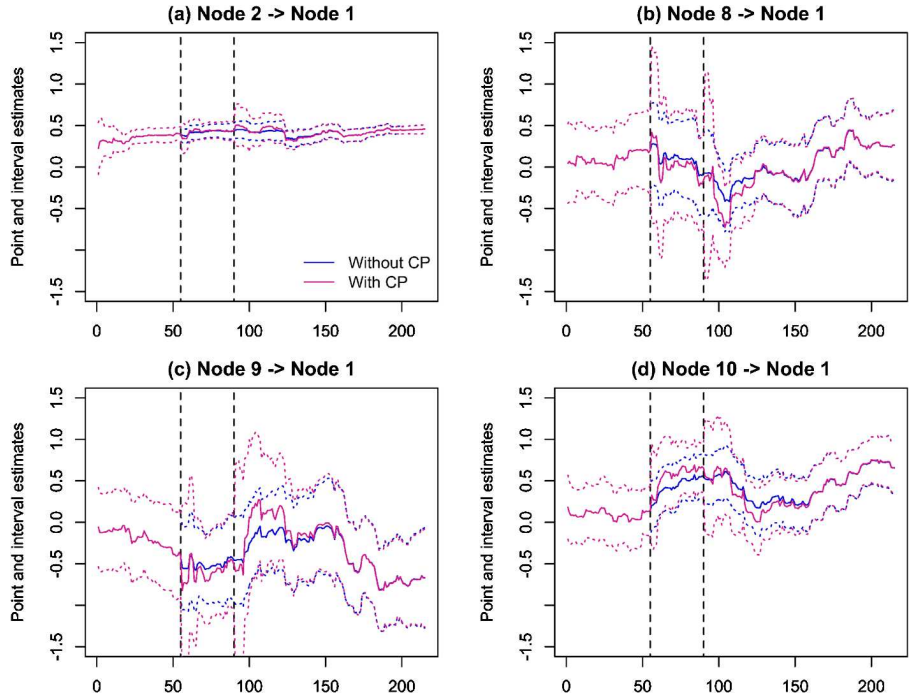}
\caption{\small{The filtering posterior mean with 95\% credible
interval for connectivities (a) Region $2 \rightarrow$ Region $1$, (b)
Region $8 \rightarrow$ Region $1$, (c) Region $9 \rightarrow$ Region
$1$ and (d) Region $10 \rightarrow$ Region $1$, considering the model
without change points (blue lines) and with change points (violet
lines). The dashed lines represent the two change points.}}
\label{ba_fig4}
\end{figure}

Adopting this method and comparing the current graph with the graph
where there is no parent from region $1$, and with a threshold of 0.3,
two time points were suggested as change points. It was straightforward
to run a new MDM with a change point at the identified point, simply by
increasing the state variance of the corresponding system error at
these two points \citep[ch.\ 11]{West}. Figure \ref{ba_fig4} shows
these two change points (dashed lines) and the filtering posterior mean
for all connectivities for this region $1$, considering both models,
without (blue lines) and with (violet lines) change points. This gives
us a different and higher scoring model, one whose score can still be
calculated in closed form.

Note that although this naive approach seems to deal adequately with
identified change points (see Figure \ref{ba_fig3}, third column), it
can obviously be improved, for example by using the full power of
switching state space models \citep[see \emph{e.g.}][ch.\
13]{fruhwirth2006finite} to model this apparent phenomenon more
formally, albeit with the loss of some simplicity. Normality and
heteroscedasticity tests were also used in this study, but neither
detected any significant deviation from the model class.

Although the iterative modifications illustrated in the application
above inevitably add additional complexity, they also allow us to
improve the model predictions and hence refine the analysis to allow
for known phenomena such as change points appearing in the signals.
Therefore they improve the selection process without entering into
complex numerical estimation methods.


\section{Conclusions}

This paper shows for the first time the use of an IP algorithm to learn
the graphical structure of MDMs.\ The conditional closure of the
associated score functions makes model selection relatively fast.\
Diagnostic statistics for checking and where necessary adapting the
whole class is also straightforward as demonstrated above. Although the
MDM was applied to resting-state fMRI data in this paper, it can be
used for other data sources, \emph{e.g.} electroencephalography. In the
same way, the MDM can also be used to estimate effective connectivity
in a task design fMRI experiment; we will show this application in
future work.

Note that we used one discount factor for each component of the MDM and
we then showed that the fitted models performed well for both synthetic
and real data. However, it is possible to specify a different discount
factor for every regression coefficient. Naturally this would be at the
expense of increasing the cost of the search analysis.

There are some exciting possibilities for using the model class to
perform an even more refined selection. For example, often the main
focus of interest in
these experiments includes not only a search for the likely model of a
specific individual, but an analysis of shared between subject effects.
Currently, such features are analysed using rather coarse aggregation
methods over shared time series. Using multivariate hierarchical models
and Bayesian
hyperclustering techniques however, it is possible to use the full
machinery of Bayesian methods to formally make inferences in a coherent
way which contemplates hypotheses about shared dependences between such
populations of subjects. In addition, diagnostic measures will be
developed for group analysis. The early results we have obtained in
building such hierarchical models are promising and again will be
reported in a later paper.

\bibliographystyle{ba}

\begin{thebibliography}{73}
\newcommand{\enquote}[1]{``#1''}

\bibitem[{Achterberg(2007)}]{Achterberg}
Achterberg, T. (2007).
\newblock\enquote{Constraint Integer Programming.}
\newblock Ph.D. thesis, Technische Universit\"at Berlin.
\endbibitem

\bibitem[{Ali et~al.(2009)Ali, Richardson, and Spirtes}]{Ali}
Ali, R.~A., Richardson, T.~S., and Spirtes, P. (2009).
\newblock\enquote{Markov equivalence for ancestral graphs.}
\newblock\emph{The Annals of Statistics\/}, 37(5B): 2808--2837.
\endbibitem

\bibitem[{Allen et~al.(2014)Allen, Damaraju, Plis, Erhardt, Eichele, and
Calhoun}]{Allen}
Allen, E.~A., Damaraju, E., Plis, S.~M., Erhardt, E.~B., Eichele, T., and
Calhoun, V.~D. (2014).
\newblock\enquote{Tracking whole-brain connectivity dynamics in the resting
state.}
\newblock\emph{Cerebral cortex\/}, 24: 663--676.
\endbibitem

\bibitem[{Anacleto et~al.(2013)Anacleto, Queen, and Albers}]{anacleto}
Anacleto, O., Queen, C., and Albers, C.~J. (2013).
\newblock\enquote{Multivariate forecasting of road traffic flows in the
presence of heteroscedasticity and measurement errors.}
\newblock\emph{Journal of the Royal Statistical Society: Series C (Applied
Statistics)\/}, 62(2): 251--270.
\endbibitem

\bibitem[{Arnhold et~al.(1999)Arnhold, Grassberger, Lehnertz, and
Elger}]{arnhold1999robust}
Arnhold, J., Grassberger, P., Lehnertz, K., and Elger, C. (1999).
\newblock\enquote{A robust method for detecting interdependences: application
to intracranially recorded EEG.}
\newblock\emph{Physica D: Nonlinear Phenomena\/}, 134(4): 419--430.
\endbibitem

\bibitem[{Baba et~al.(2004)Baba, Shibata, and Sibuya}]{baba2004partial}
Baba, K., Shibata, R., and Sibuya, M. (2004).
\newblock\enquote{Partial correlation and conditional correlation as measures
of conditional independence.}
\newblock\emph{Australian \& New Zealand Journal of Statistics\/}, 46(4):
657--664.
\endbibitem

\bibitem[{Bartlett and Cussens(2013)}]{bartlett13}
Bartlett, M. and Cussens, J. (2013).
\newblock\enquote{Advances in Bayesian Network Learning Using Integer
Programming.}
\newblock In \emph{Proceedings of the 29th Conference on Uncertainty in
Artificial Intelligence (UAI 2013)\/}, 182--191. AUAI Press.
\endbibitem

\bibitem[{Benjamini and Hochberg(1995)}]{benjamini1995controlling}
Benjamini, Y. and Hochberg, Y. (1995).
\newblock\enquote{Controlling the false discovery rate: a practical and
powerful approach to multiple testing.}
\newblock\emph{Journal of the Royal Statistical Society. Series B
(Methodological)\/}, 289--300.
\endbibitem

\bibitem[{Bhattacharya and Maitra(2011)}]{bhattacharya2011nonstationary}
Bhattacharya, S. and Maitra, R. (2011).
\newblock\enquote{A nonstationary nonparametric Bayesian approach to
dynamically modeling effective connectivity in functional magnetic resonance
imaging experiments.}
\newblock\emph{The Annals of Applied Statistics\/}, 5(2B): 1183--1206.
\endbibitem

\bibitem[{Bhattacharya et~al.(2006)Bhattacharya, Ringo~Ho, and
Purkayastha}]{bhattacharya2006bayesian}
Bhattacharya, S., Ringo~Ho, M.-H., and Purkayastha, S. (2006).
\newblock\enquote{A Bayesian approach to modeling dynamic effective
connectivity with fMRI data.}
\newblock\emph{Neuroimage\/}, 30(3): 794--812.
\endbibitem

\bibitem[{Chang and Glover(2010)}]{chang2010time}
Chang, C. and Glover, G.~H. (2010).
\newblock\enquote{Time--frequency dynamics of resting-state brain connectivity
measured with fMRI.}
\newblock\emph{Neuroimage\/}, 50(1): 81--98.
\endbibitem

\bibitem[{Chang et~al.(2008)Chang, Thomason, and Glover}]{chang2008mapping}
Chang, C., Thomason, M.~E., and Glover, G.~H. (2008).
\newblock\enquote{Mapping and correction of vascular hemodynamic
latency in
the BOLD signal.}
\newblock\emph{Neuroimage\/}, 43(1): 90--102.
\endbibitem

\bibitem[{Chickering(2003)}]{chickering2003optimal}
Chickering, D.~M. (2003).
\newblock\enquote{Optimal structure identification with greedy search.}
\newblock\emph{The Journal of Machine Learning Research\/}, 3: 507--554.
\endbibitem

\bibitem[{Consonni and La~Rocca(2011)}]{Consonni}
Consonni, G. and La~Rocca, L. (2011).
\newblock\enquote{Moment priors for Bayesian model choice with
applications to
directed acyclic graphs.}
\newblock\emph{Bayesian Statistics 9\/}, 9: 119.
\endbibitem

\bibitem[{Cowell(2013)}]{cowell2013simple}
Cowell, R.~G. (2013).
\newblock\enquote{A simple greedy algorithm for reconstructing pedigrees.}
\newblock\emph{Theoretical Population Biology\/}, 83: 55--63.
\endbibitem

\bibitem[{Cowell et~al.(1999)Cowell, Dawid, Lauritzen, and
Spiegelhalter}]{Cowell99}
Cowell, R.~G., Dawid, A.~P., Lauritzen, S.~L., and Spiegelhalter, D.~J. (1999).
\newblock\emph{Probabilistic Networks and Expert Systems\/}.
\newblock New York: Springer-Verlag.
\endbibitem

\bibitem[{Cribben et~al.(2012)Cribben, Haraldsdottir, Atlas, Wager, and
Lindquist}]{cribben2012dynamic}
Cribben, I., Haraldsdottir, R., Atlas, L.~Y., Wager, T.~D., and Lindquist,
M.~A. (2012).
\newblock\enquote{Dynamic connectivity regression: determining state-related
changes in brain connectivity.}
\newblock\emph{Neuroimage\/}, 61(4): 907--920.
\endbibitem

\bibitem[{Cussens(2010)}]{Cussens10}
Cussens, J. (2010).
\newblock\enquote{Maximum likelihood pedigree reconstruction using integer
programming.}
\newblock In \emph{Proceedings of the Workshop on Constraint Based
Methods for
Bioinformatics (WCB-10)\/}. Edinburgh.
\endbibitem

\bibitem[{Cussens(2011)}]{cussens11}
--- (2011).
\newblock\enquote{Bayesian Network Learning with Cutting Planes.}
\newblock In Cozman, F.~G. and Pfeffer, A. (eds.), \emph{Proceedings of
the 27th
Conference on Uncertainty in Artificial Intelligence (UAI 2011)\/}, 153--160.
Barcelona: AUAI Press.
\newline URL \url{http://uai.sis.pitt.edu/papers/11/p153-cussens.pdf}
\endbibitem

\bibitem[{Daunizeau et~al.(2009)Daunizeau, Friston, and
Kiebel}]{daunizeau2009variational}
Daunizeau, J., Friston, K., and Kiebel, S. (2009).
\newblock\enquote{Variational Bayesian identification and prediction of
stochastic nonlinear dynamic causal models.}
\newblock\emph{Physica D: Nonlinear Phenomena\/}, 238(21): 2089--2118.
\endbibitem

\bibitem[{Dauwels et~al.(2010)Dauwels, Vialatte, Musha, and
Cichocki}]{dauwels2010comparative}
Dauwels, J., Vialatte, F., Musha, T., and Cichocki, A. (2010).
\newblock\enquote{A comparative study of synchrony measures for the early
diagnosis of Alzheimer's disease based on EEG.}
\newblock\emph{NeuroImage\/}, 49(1): 668--693.
\endbibitem

\bibitem[{David et~al.(2008)David, Guillemain, Saillet, Reyt, Deransart,
Segebarth, and Depaulis}]{david2008identifying}
David, O., Guillemain, I., Saillet, S., Reyt, S., Deransart, C.,
Segebarth, C.,
and Depaulis, A. (2008).
\newblock\enquote{Identifying neural drivers with functional MRI: an
electrophysiological validation.}
\newblock\emph{PLoS Biology\/}, 6(12): e315.
\endbibitem

\bibitem[{Duff et~al.(2013)Duff, Makin, Madugula, Smith, and
Woolrich}]{duff2013utility}
Duff, E., Makin, T., Madugula, S., Smith, S.~M., and Woolrich, M.~W. (2013).
\newblock\enquote{Utility of Partial Correlation for Characterising Brain
Dynamics: MVPA-based Assessment of Regularisation and Network Selection.}
\newblock In \emph{Pattern Recognition in Neuroimaging (PRNI), 2013
International Workshop on\/}, 58--61. IEEE.
\endbibitem

\bibitem[{Durbin and Koopman(2012)}]{durbin2012time}
Durbin, J. and Koopman, S.~J. (2012).
\newblock\emph{Time series analysis by state space methods\/}.
\newblock38. Oxford University Press.
\endbibitem

\bibitem[{Friston(2011)}]{friston2011functional}
Friston, K.~J. (2011).
\newblock\enquote{Functional and effective connectivity: a review.}
\newblock\emph{Brain Connectivity\/}, 1(1): 13--36.
\endbibitem

\bibitem[{Friston et~al.(2003)Friston, Harrison, and
Penny}]{friston2003dynamic}
Friston, K.~J., Harrison, L., and Penny, W. (2003).
\newblock\enquote{Dynamic causal modelling.}
\newblock\emph{Neuroimage\/}, 19(4): 1273--1302.
\endbibitem

\bibitem[{Fr{\"u}hwirth-Schnatter(2006)}]{fruhwirth2006finite}
Fr{\"u}hwirth-Schnatter, S. (2006).
\newblock\emph{Finite Mixture and Markov Switching Models: Modeling and
Applications to Random Processes\/}.
\newblock Springer.
\endbibitem

\bibitem[{Ge et~al.(2009)Ge, Kendrick, and Feng}]{ge2009novel}
Ge, T., Kendrick, K.~M., and Feng, J. (2009).
\newblock\enquote{A novel extended Granger causal model approach demonstrates
brain hemispheric differences during face recognition learning.}
\newblock\emph{PLoS computational biology\/}, 5(11): e1000570.
\endbibitem

\bibitem[{Granger(1969)}]{granger1969investigating}
Granger, C.~W. (1969).
\newblock\enquote{Investigating causal relations by econometric models and
cross-spectral methods.}
\newblock\emph{Econometrica: Journal of the Econometric Society\/}, 37(3):
424--438.
\endbibitem

\bibitem[{Harrison and West(1991)}]{harrison1991dynamic}
Harrison, J. and West, M. (1991).
\newblock\enquote{Dynamic linear model diagnostics.}
\newblock\emph{Biometrika\/}, 78(4): 797--808.
\endbibitem

\bibitem[{Havlicek et~al.(2010)Havlicek, Jan, Brazdil, and
Calhoun}]{havlicek2010dynamic}
Havlicek, M., Jan, J., Brazdil, M., and Calhoun, V.~D. (2010).
\newblock\enquote{Dynamic Granger causality based on Kalman filter for
evaluation of functional network connectivity in fMRI data.}
\newblock\emph{Neuroimage\/}, 53(1): 65--77.
\endbibitem

\bibitem[{Heard et~al.(2006)Heard, Holmes, and
Stephens}]{heard2006quantitative}
Heard, N.~A., Holmes, C.~C., and Stephens, D.~A. (2006).
\newblock\enquote{A quantitative study of gene regulation involved in the
immune response of anopheline mosquitoes: An application of Bayesian
hierarchical clustering of curves.}
\newblock\emph{Journal of the American Statistical Association\/}, 101(473):
18--29.
\endbibitem

\bibitem[{Heckerman(1998)}]{heckerman1998tutorial}
Heckerman, D. (1998).
\newblock\emph{A tutorial on learning with Bayesian networks\/}.
\newblock Springer.
\endbibitem

\bibitem[{Jaakkola et~al.(2010)Jaakkola, Sontag, Globerson, and
Meila}]{jaakkola10}
Jaakkola, T., Sontag, D., Globerson, A., and Meila, M. (2010).
\newblock\enquote{Learning Bayesian Network Structure using {LP} Relaxations.}
\newblock In \emph{Proceedings of 13th International Conference on Artificial
Intelligence and Statistics (AISTATS 2010)\/}, volume~9, 358--365.
\newblock Journal of Machine Learning Research Workshop and Conference
Proceedings.
\endbibitem

\bibitem[{Jeffreys(1998)}]{jeffreys1998theory}
Jeffreys, H. (1998).
\newblock\emph{The theory of probability\/}.
\newblock Oxford University Press.
\endbibitem

\bibitem[{Kalisch and B{\"u}hlmann(2008)}]{kalisch2008robustification}
Kalisch, M. and B{\"u}hlmann, P. (2008).
\newblock\enquote{Robustification of the PC-algorithm for Directed Acyclic
Graphs.}
\newblock\emph{Journal of Computational and Graphical Statistics\/}, 17(4):
773--789.
\endbibitem

\bibitem[{Korb and Nicholson(2003)}]{korb2003bayesian}
Korb, K.~B. and Nicholson, A.~E. (2003).
\newblock\emph{Bayesian artificial intelligence\/}.
\newblock cRc Press.
\endbibitem

\bibitem[{Lauritzen(1996)}]{lauritzen1996graphical}
Lauritzen, S.~L. (1996).
\newblock\emph{Graphical models\/}.
\newblock Oxford University Press.
\endbibitem

\bibitem[{Leonardi et~al.(2013)Leonardi, Richiardi, Gschwind, Simioni, Annoni,
Schluep, Vuilleumier, and Van De~Ville}]{leonardi2013principal}
Leonardi, N., Richiardi, J., Gschwind, M., Simioni, S., Annoni, J.-M., Schluep,
M., Vuilleumier, P., and Van De~Ville, D. (2013).
\newblock\enquote{Principal components of functional connectivity: a new
approach to study dynamic brain connectivity during rest.}
\newblock\emph{NeuroImage\/}, 83: 937--950.
\endbibitem

\bibitem[{Li et~al.(2011)Li, Daunizeau, Stephan, Penny, Hu, and
Friston}]{li2011generalised}
Li, B., Daunizeau, J., Stephan, K.~E., Penny, W., Hu, D., and Friston, K.
(2011).
\newblock\enquote{Generalised filtering and stochastic DCM for fMRI.}
\newblock\emph{Neuroimage\/}, 58(2): 442--457.
\endbibitem

\bibitem[{Marrelec et~al.(2006)Marrelec, Krainik, Duffau,
P{\'e}l{\'e}grini-Issac, Leh{\'e}ricy, Doyon, and
Benali}]{marrelec2006partial}
Marrelec, G., Krainik, A., Duffau, H., P{\'e}l{\'e}grini-Issac, M.,
Leh{\'e}ricy, S., Doyon, J., and Benali, H. (2006).
\newblock\enquote{Partial correlation for functional brain interactivity
investigation in functional MRI.}
\newblock\emph{Neuroimage\/}, 32(1): 228--237.
\endbibitem

\bibitem[{Meek(1995)}]{meek1995causal}
Meek, C. (1995).
\newblock\enquote{Causal inference and causal explanation with background
knowledge.}
\newblock In \emph{Proceedings of the Eleventh Conference on Uncertainty in
Artificial Intelligence\/}, 403--410. Morgan Kaufmann Publishers Inc.
\endbibitem

\bibitem[{Meek(1997)}]{Meek97}
--- (1997).
\newblock\enquote{Graphical Models: Selecting causal and statistical models.}
\newblock Ph.D. thesis, Carnegie Mellon University.
\endbibitem

\bibitem[{Patel et~al.(2006)Patel, Bowman, and Rilling}]{patel2006bayesian}
Patel, R.~S., Bowman, F.~D., and Rilling, J.~K. (2006).
\newblock\enquote{A Bayesian approach to determining connectivity of
the human
brain.}
\newblock\emph{Human brain mapping\/}, 27(3): 267--276.
\endbibitem

\bibitem[{Pearl(2000)}]{pearl2000causality}
Pearl, J. (2000).
\newblock\emph{Causality: models, reasoning and inference\/}, volume~29.
\newblock Cambridge University Press.
\endbibitem

\bibitem[{Pearl(2009)}]{pearl2009causal}
--- (2009).
\newblock\enquote{Causal inference in statistics: An overview.}
\newblock\emph{Statistics Surveys\/}, 3: 96--146.
\endbibitem

\bibitem[{Penny et~al.(2005)Penny, Ghahramani, and Friston}]{penny2005bilinear}
Penny, W., Ghahramani, Z., and Friston, K. (2005).
\newblock\enquote{Bilinear dynamical systems.}
\newblock\emph{Philosophical Transactions of the Royal Society B: Biological
Sciences\/}, 360(1457): 983--993.
\endbibitem

\bibitem[{Pereda et~al.(2005)Pereda, Quiroga, and
Bhattacharya}]{pereda2005nonlinear}
Pereda, E., Quiroga, R.~Q., and Bhattacharya, J. (2005).
\newblock\enquote{Nonlinear multivariate analysis of neurophysiological
signals.}
\newblock\emph{Progress in neurobiology\/}, 77(1): 1--37.
\endbibitem

\bibitem[{Petris et~al.(2009)Petris, Petrone, and
Campagnoli}]{petris2009dynamic}
Petris, G., Petrone, S., and Campagnoli, P. (2009).
\newblock\emph{Dynamic linear models with R\/}.
\newblock Springer.
\endbibitem

\bibitem[{Poldrack et~al.(2011)Poldrack, Mumford, and
Nichols}]{poldrack2011handbook}
Poldrack, R.~A., Mumford, J.~A., and Nichols, T.~E. (2011).
\newblock\emph{Handbook of functional MRI data analysis\/}.
\newblock Cambridge University Press.
\endbibitem

\bibitem[{Queen and Albers(2009)}]{queen2009intervention}
Queen, C.~M. and Albers, C.~J. (2009).
\newblock\enquote{Intervention and causality: forecasting traffic
flows using
a dynamic Bayesian network.}
\newblock\emph{Journal of the American Statistical Association\/}, 104(486):
669--681.
\endbibitem

\bibitem[{Queen and Smith(1993)}]{queen1993multiregression}
Queen, C.~M. and Smith, J.~Q. (1993).
\newblock\enquote{Multiregression dynamic models.}
\newblock\emph{Journal of the Royal Statistical Society. Series B
(Methodological)\/}, 55(4): 849--870.
\endbibitem

\bibitem[{Queen et~al.(2008)Queen, Wright, and Albers}]{queen2008forecast}
Queen, C.~M., Wright, B.~J., and Albers, C.~J. (2008).
\newblock\enquote{Forecast covariances in the linear multiregression dynamic
model.}
\newblock\emph{Journal of Forecasting\/}, 27(2): 175--191.
\endbibitem

\bibitem[{Quian~Quiroga et~al.(2002)Quian~Quiroga, Kraskov, Kreuz, and
Grassberger}]{quiroga2002performance}
Quian~Quiroga, R., Kraskov, A., Kreuz, T., and Grassberger, P. (2002).
\newblock\enquote{Performance of different synchronization measures in real
data: a case study on electroencephalographic signals.}
\newblock\emph{Physical Review E\/}, 65(4): 041903.
\endbibitem

\bibitem[{Ramsey et~al.(2010)Ramsey, Hanson, Hanson, Halchenko,
Poldrack, and
Glymour}]{ramsey2010six}
Ramsey, J.~D., Hanson, S.~J., Hanson, C., Halchenko, Y.~O., Poldrack, R.~A.,
and Glymour, C. (2010).
\newblock\enquote{Six problems for causal inference from fMRI.}
\newblock\emph{Neuroimage\/}, 49(2): 1545--1558.
\endbibitem

\bibitem[{Roebroeck et~al.(2011)Roebroeck, Formisano, and
Goebel}]{roebroeck2011identification}
Roebroeck, A., Formisano, E., and Goebel, R. (2011).
\newblock\enquote{The identification of interacting networks in the brain
using fMRI: model selection, causality and deconvolution.}
\newblock\emph{Neuroimage\/}, 58(2): 296--302.
\endbibitem

\bibitem[{Ryali et~al.(2011)Ryali, Supekar, Chen, and
Menon}]{ryali2011multivariate}
Ryali, S., Supekar, K., Chen, T., and Menon, V. (2011).
\newblock\enquote{Multivariate dynamical systems models for estimating causal
interactions in fMRI.}
\newblock\emph{Neuroimage\/}, 54(2): 807--823.
\endbibitem

\bibitem[{Salimi-Khorshidi et~al.(2014)Salimi-Khorshidi, Douaud, Beckmann,
Glasser, Griffanti, and Smith}]{salimi2014automatic}
Salimi-Khorshidi, G., Douaud, G., Beckmann, C.~F., Glasser, M.~F., Griffanti,
L., and Smith, S.~M. (2014).
\newblock\enquote{Automatic denoising of functional MRI data: combining
independent component analysis and hierarchical fusion of classifiers.}
\newblock\emph{NeuroImage\/}, 90: 449--468.
\endbibitem

\bibitem[{Schwarz et~al.(1978)}]{schwarz1978estimating}
Schwarz, G. et~al. (1978).
\newblock\enquote{Estimating the dimension of a model.}
\newblock\emph{The annals of statistics\/}, 6(2): 461--464.
\endbibitem

\bibitem[{Shimizu et~al.(2006)Shimizu, Hoyer, Hyv{\"a}rinen, and
Kerminen}]{shimizu2006linear}
Shimizu, S., Hoyer, P.~O., Hyv{\"a}rinen, A., and Kerminen, A. (2006).
\newblock\enquote{A linear non-Gaussian acyclic model for causal discovery.}
\newblock\emph{The Journal of Machine Learning Research\/}, 7: 2003--2030.
\endbibitem

\bibitem[{Smith(1985)}]{smith1985diagnostic}
Smith, J. (1985).
\newblock\enquote{Diagnostic checks of non-standard time series models.}
\newblock\emph{Journal of Forecasting\/}, 4(3): 283--291.
\endbibitem

\bibitem[{Smith et~al.(2010)Smith, Pillai, Chen, and
Horwitz}]{smith2010identification}
Smith, J.~F., Pillai, A., Chen, K., and Horwitz, B. (2010).
\newblock\enquote{Identification and validation of effective connectivity
networks in functional magnetic resonance imaging using switching linear
dynamic systems.}
\newblock\emph{Neuroimage\/}, 52(3): 1027--1040.
\endbibitem

\bibitem[{Smith et~al.(2011{\natexlab{a}})Smith, Pillai, Chen, and
Horwitz}]{smith2011effective}
--- (2011{\natexlab{a}}).
\newblock\enquote{Effective connectivity modeling for fMRI: six issues and
possible solutions using linear dynamic systems.}
\newblock\emph{Frontiers in systems neuroscience\/}, 5(104).
\endbibitem

\bibitem[{Smith et~al.(2012)Smith, Bandettini, Miller, Behrens,
Friston, David,
Liu, Woolrich, and Nichols}]{smith2012danger}
Smith, S.~M., Bandettini, P.~A., Miller, K.~L., Behrens, T., Friston, K.~J.,
David, O., Liu, T., Woolrich, M.~W., and Nichols, T.~E. (2012).
\newblock\enquote{The danger of systematic bias in group-level FMRI-lag-based
causality estimation.}
\newblock\emph{Neuroimage\/}, 59(2): 1228--1229.
\endbibitem

\bibitem[{Smith et~al.(2011{\natexlab{b}})Smith, Miller, Salimi-Khorshidi,
Webster, Beckmann, Nichols, Ramsey, and Woolrich}]{smith2011network}
Smith, S.~M., Miller, K.~L., Salimi-Khorshidi, G., Webster, M., Beckmann,
C.~F., Nichols, T.~E., Ramsey, J.~D., and Woolrich, M.~W.
(2011{\natexlab{b}}).
\newblock\enquote{Network modelling methods for FMRI.}
\newblock\emph{Neuroimage\/}, 54(2): 875--891.
\endbibitem

\bibitem[{Spirtes et~al.(2000)Spirtes, Glymour, and
Scheines}]{spirtes2000causation}
Spirtes, P., Glymour, C.~N., and Scheines, R. (2000).
\newblock\emph{Causation, prediction, and search\/}, volume~81.
\newblock MIT press.
\endbibitem

\bibitem[{Steinsky(2003)}]{steinsky2003enumeration}
Steinsky, B. (2003).
\newblock\enquote{Enumeration of labelled chain graphs and labelled essential
directed acyclic graphs.}
\newblock\emph{Discrete Mathematics\/}, 270(1): 267--278.
\endbibitem

\bibitem[{Stephan et~al.(2008)Stephan, Kasper, Harrison, Daunizeau, den Ouden,
Breakspear, and Friston}]{stephan2008nonlinear}
Stephan, K.~E., Kasper, L., Harrison, L.~M., Daunizeau, J., den Ouden, H.~E.,
Breakspear, M., and Friston, K.~J. (2008).
\newblock\enquote{Nonlinear dynamic causal models for fMRI.}
\newblock\emph{Neuroimage\/}, 42(2): 649--662.
\endbibitem

\bibitem[{Stephan et~al.(2010)Stephan, Penny, Moran, den Ouden,
Daunizeau, and
Friston}]{stephan2010ten}
Stephan, K.~E., Penny, W.~D., Moran, R.~J., den Ouden, H.~E.,
Daunizeau, J.,
and Friston, K.~J. (2010).
\newblock\enquote{Ten simple rules for dynamic causal modeling.}
\newblock\emph{Neuroimage\/}, 49(4): 3099--3109.
\endbibitem

\bibitem[{Vald\'es-Sosa et~al.(2011)Vald\'es-Sosa, Roebroeck,
Daunizeau, and
Friston}]{valdes2011effective}
Vald\'es-Sosa, P.~A., Roebroeck, A., Daunizeau, J., and Friston, K. (2011).
\newblock\enquote{Effective connectivity: influence, causality and biophysical
modeling.}
\newblock\emph{Neuroimage\/}, 58(2): 339--361.
\endbibitem

\bibitem[{West and Harrison(1997)}]{West}
West, M. and Harrison, P.~J. (1997).
\newblock\emph{Bayesian Forecasting and Dynamic Models\/}.
\newblock New York: Springer-Verlag, 2nd edition.
\endbibitem

\bibitem[{Wolsey(1998)}]{Wolsey}
Wolsey, L.~A. (1998).
\newblock\emph{Integer Programming\/}.
\newblock John Wiley.
\endbibitem

\bibitem[{Zhang et~al.(2014)Zhang, Li, Li, Lian, Huang, Zhong, Zhu, Li,
Jin, Hu
et~al.}]{zhang2013inferring}
Zhang, J., Li, X., Li, C., Lian, Z., Huang, X., Zhong, G., Zhu, D., Li, K.,
Jin, C., Hu, X., et~al. (2014).
\newblock\enquote{Inferring functional interaction and transition
patterns via
dynamic Bayesian variable partition models.}
\newblock\emph{Human brain mapping\/}, 35: 3314--3331.
\endbibitem

\end{thebibliography}

%
\begin{acknowledgement}
This work has been supported by the UK Medical Research Council
(Project Grant G1002312) and by CAPES (Coordena\c{c}\~ao de Aperfei\c
{c}oamento de Pessoal de N\'ivel Superior), Brazil.
\end{acknowledgement}


\appendix
\section*{Appendix A}

The graphical structure used to obtain the synthetic data discussed in
Section 4 is shown in Figure \ref{ba_fig1a} (b). The initial values for
the regression parameters were defined as the average of estimated
values over time from the real data, \emph{i.e.} zero for intercept
parameters, $0.25$ for connection $Y(2) \rightarrow Y(4)$, $0.18$ for
connection $Y(8) \rightarrow Y(4)$, $0.50$ for connection $Y(3)
\rightarrow Y(5)$, $0.80$ for connection $Y(7) \rightarrow Y(6)$,
$0.39$ for connection $Y(8) \rightarrow Y(7)$, and $0.65$ for
connection $Y(10) \rightarrow Y(9)$. The observational variance was
also defined considering the estimated variance of variables from the
real data, \emph{i.e.} $0.010$, $0.191$, $0.036$, $0.005$, $0.018$,
$0.011$, $0.010$, $0.006$, $0.016$, $0.014$ and $0.013$ for the
variables of nodes $1$ to $11$, respectively. Thus we set
\begin{eqnarray*}
\theta_{ti}^{(k)}(r) = \theta_{t-1i}^{(k)}(r) + w_{ti}^{(k)}(r) \text{,
~~~} w_{ti}^{(k)}(r) \sim\mathcal{N} (0, W^{(k)}(r)),
\end{eqnarray*}
for $r=1,\ldots,11$; $t=1, \dots, 230$ ; $i=1, \dots, 50$ replications
(the same as the last section); $k=1, \dots, p_r$; $p_r = 1$, for $r
\in\{1,2,3,8,10,11\}$; $p_r = 2$, for $r \in\{5,6,7,9\}$; $p_4 = 2$;
$W^{(k)}(r)=W^{*(k)}(r) \times V(r)$ and $W^{*(k)}(r)$ is the $k^{th}$
element of the diagonal of matrix $\mathbf{W}^*(r) = 0.05\mathbf
{I}_{p_r}$. Observed values were then simulated using the following equations:
\begin{eqnarray*}
Y_{ti}(j) &=& \theta_{ti}^{(1)}(j) + v_{ti}(j);\\
Y_{ti}(4) &=& \theta_{ti}^{(1)}(4) + \theta_{ti}^{(2)}(4)Y_{ti}(2) +
\theta_{ti}^{(3)}(4)Y_{ti}(8) + v_{ti}(4);\\
Y_{ti}(5) &=& \theta_{ti}^{(1)}(5) + \theta_{ti}^{(2)}(5)Y_{ti}(3) +
v_{ti}(5);\\
Y_{ti}(7) &=& \theta_{ti}^{(1)}(7) + \theta_{ti}^{(2)}(7)Y_{ti}(8) +
v_{ti}(7);\\
Y_{ti}(6) &=& \theta_{ti}^{(1)}(6) + \theta_{ti}^{(2)}(6)Y_{ti}(7) +
v_{ti}(6);\\
Y_{ti}(9) &=& \theta_{ti}^{(1)}(9) + \theta_{ti}^{(2)}(9)Y_{ti}(10) +
v_{ti}(9);
\end{eqnarray*}
where $j \in\{1,2,3,8,10,11\}$, $v_{ti}(r) \sim\mathcal{N}(0,V(r))$,
and other parameters were defined as before.

\section*{Appendix B}

The graphical structure used to obtain the synthetic data discussed in
Section 5 is shown in Figure \ref{fig5ni} (a) DAG1. The initial values
for the regression parameters were $0.3$ for the connectivity between
$Y(1)$ and $Y(2)$, \emph{i.e.} $\theta_0^{(2)}(2)$, $0.2$ for the
connectivity between $Y(2)$ and $Y(3)$, \emph{i.e.} $\theta
_0^{(2)}(3)$, and the value $0$ for other $\theta$'s (intercept
parameters). The observational variance was defined as $12.5$ for
$Y(1)$, $6.3$ for $Y(2)$ and $5.0$ for $Y(3)$, so that the marginal
variances were almost the same for both regions. Thus we set
\begin{eqnarray*}
\theta_{ti}^{(k)}(r) = \theta_{t-1i}^{(k)}(r) + w_{ti}^{(k)}(r) \text{,
~~~} w_{ti}^{(k)}(r) \sim\mathcal{N} (0, W^{(k)}(r)),
\end{eqnarray*}
for $r=1, \ldots, 3$; $t=1, \dots, T$ ; $i=1, \dots, 100$ replications;
$k=1, \dots, p_r$; $p_1 = 1$; $p_2 = 2$; $p_3=2$;
$W^{(k)}(r)=W^{*(k)}(r) \times V(r)$ and $W^{*(k)}(r)$ is the $k^{th}$
element of the diagonal of matrix $\mathbf{W}^*(r)$ defined above.
Observed values were then simulated using the following equations:
\begin{eqnarray*}
Y_{ti}(1) &=& \theta_{ti}^{(1)}(1) + v_{ti}(1) \text{, ~~~~~} v_{ti}(1)
\sim\mathcal{N}(0,V(1));\\
Y_{ti}(2) &=& \theta_{ti}^{(1)}(2) + \theta_{ti}^{(2)}(2)Y_{ti}(1) +
v_{ti}(2) \text{, ~~~~~} v_{ti}(2) \sim\mathcal{N}(0,V(2));\\
Y_{ti}(3) &=& \theta_{ti}^{(1)}(3) + \theta_{ti}^{(2)}(3)Y_{ti}(2) +
v_{ti}(3) \text{, ~~~~~} v_{ti}(3) \sim\mathcal{N}(0,V(3)).
\end{eqnarray*}

\end{document}